\documentclass[pra, notitlepage, showpacs, floatfix, superscriptaddress, reprint]{revtex4-1}
\usepackage{amsmath}
\usepackage{amssymb}
\usepackage{amsthm}
\usepackage{amsfonts}
\usepackage{verbatim}
\usepackage{listings}
\usepackage{enumerate}
\usepackage{latexsym}
\usepackage{psfrag}
\usepackage{bm}
\usepackage[all]{xy}
\usepackage{graphicx}
\usepackage{color}
\usepackage{mathtools}
\usepackage{eufrak}
\usepackage[percent]{overpic}
\usepackage{epsfig,slashed}
\usepackage{epstopdf}
\usepackage{lipsum}
\usepackage{float}
\usepackage{mathtools}
\usepackage[colorlinks=true,citecolor=blue,linkcolor=black,urlcolor=blue]{hyperref}
\usepackage{natbib}
\usepackage{subfigure}
\usepackage[mathscr]{euscript}
\usepackage[export]{adjustbox}
\usepackage[bottom]{footmisc}
\usepackage{textcomp}
\usepackage{braket}
\makeatletter
\newsavebox\myboxA
\newsavebox\myboxB
\newlength\mylenA

\newcommand*\xoverline[2][0.75]{%
    \sbox{\myboxA}{$\m@th#2$}%
    \setbox\myboxB\null% Phantom box
    \ht\myboxB=\ht\myboxA%
    \dp\myboxB=\dp\myboxA%
    \wd\myboxB=#1\wd\myboxA% Scale phantom
    \sbox\myboxB{$\m@th\overline{\copy\myboxB}$}%  Overlined phantom
    \setlength\mylenA{\the\wd\myboxA}%   calc width diff
    \addtolength\mylenA{-\the\wd\myboxB}%
    \ifdim\wd\myboxB<\wd\myboxA%
       \rlap{\hskip 0.5\mylenA\usebox\myboxB}{\usebox\myboxA}%
    \else
        \hskip -0.5\mylenA\rlap{\usebox\myboxA}{\hskip 0.5\mylenA\usebox\myboxB}%
    \fi}
\makeatother

\def\II{\hbox{$1\hskip -1.2pt\vrule depth 0pt height 1.6ex width 0.7pt\vrule depth 0pt height 0.3pt width 0.12em$}}
\DeclareGraphicsExtensions{.png}
\begin{document}

\title{Numerical study of the chiral $\mathbb{Z}_3$ quantum phase transition in one spatial dimension}
\author{Rhine Samajdar}
\affiliation{$\mbox{Department of Physics, Harvard University, Cambridge, MA 02138, USA}$}
\author{Soonwon Choi}
\affiliation{$\mbox{Department of Physics, Harvard University, Cambridge, MA 02138, USA}$}
\author{Hannes Pichler}
\affiliation{$\mbox{Department of Physics, Harvard University, Cambridge, MA 02138, USA}$}
\affiliation{$\mbox{ITAMP, Harvard-Smithsonian Center for Astrophysics, Cambridge, MA 02138, USA}$}
\author{Mikhail D. Lukin}
\affiliation{$\mbox{Department of Physics, Harvard University, Cambridge, MA 02138, USA}$}
\author{Subir Sachdev}
\affiliation{$\mbox{Department of Physics, Harvard University, Cambridge, MA 02138, USA}$}
\affiliation{$\mbox{Perimeter Institute for Theoretical Physics, Waterloo, Ontario N2L 2Y5, Canada}$}

\begin{abstract}
Recent experiments on a one-dimensional chain of trapped 
alkali atoms [Bernien {\it et al.}, Nature {\bf 551}, 579 (2017)] have observed a quantum transition associated with the onset of period-3 ordering of pumped Rydberg states. This spontaneous $\mathbb{Z}_3$ symmetry breaking is described by a constrained model of hard-core bosons proposed by Fendley {\it et al.} [Phys. Rev. B {\bf 69}, 075106 (2004)]. By symmetry arguments, the transition is expected to be in the universality class of the $\mathbb{Z}_3$ chiral clock model with parameters preserving both time-reversal and spatial-inversion symmetries. We study the nature of the order--disorder transition in these models, and numerically calculate its critical exponents with exact diagonalization and density-matrix renormalization group techniques. We use finite-size scaling to determine the dynamical critical exponent $z$ and the correlation length exponent $\nu$. Our analysis presents the only known instance of a strongly coupled generic transition between gapped states with $z \ne 1$, implying an underlying nonconformal critical field theory.
\end{abstract}
\maketitle

\hypersetup{linkcolor=blue}

\section{Introduction}
\label{sec:intro}
In recent years, symmetry-breaking quantum phase transitions (QPTs) of bosons and spin systems have been extensively studied \cite{sachdev2011quantum}, and many experimental realizations have been found \cite{bitko1996quantum,greiner2002quantum,ruegg2008quantum,coldea2010quantum}. 
In all the noted cases, the zero-temperature quantum critical properties of the transitions are largely well understood.

A new realization of QPTs was recently found by \citet{bernien2017probing} in a one-dimensional chain of trapped alkali atoms, where they observed the onset of modulation in pumped Rydberg states. A lattice model of hard-core bosons \cite{fendley2004competing} provides a good description 
of the experiments. Both the experiments and the theoretical model display a QPT with period-3 ordering. When combined with the requirements of time-reversal and spatial-inversion symmetries, the period-3 ordering implies that the QPT should be in the universality class
of the $\mathbb{Z}_3$ chiral clock model \cite{huse1981simple,ostlund1981incommensurate,huse1982domain,huse1983melting,haldane1983phase,howes1983quantum,au1987commuting,fendley2012parafermionic,zhuang2015phase} over a set of parameters ($\phi=0$, $\theta\neq 0$), which will be specified below. Curiously, while this clock model has been studied theoretically for over three decades, there is no controlled field-theoretic description for a QPT with the specified symmetries between a gapped $\mathbb{Z}_3$-ordered
state, and a gapped disordered state that preserves translational symmetry. Our paper will present numerical results which shed light on the nature of this novel strongly coupled QPT in 1+1 dimensions.

One of the reasons for the tractability of the previously studied QPTs is that they all have dynamical critical exponent $z=1$. Indeed, their critical theories
are relativistically and conformally invariant, and this large symmetry enables much analytic progress. Our results here clearly show 
that the chiral clock transition has dynamical critical exponent $z \neq 1$. To our knowledge, this transition provides the only known
strongly coupled critical point for a generic QPT between gapped states which has $z \neq 1$, and so, cannot be described by a conformal field theory. 
A prior example \cite{SS96,KY04,Sau17} of a strongly coupled QPT with $z \neq 1$ involved the the onset of Ising ferromagnetic order in a one-dimensional quantum spin system with $U(1)$ spin-rotation symmetry about the Ising axis; however, in this case, the phases flanking the critical point are both gapless. Nongeneric QPTs with $z\neq 1$, and Hamiltonians which are finely tuned to obtain a known ground-state wavefunction, appear in Refs.~\onlinecite{VBS04,FHMOS04,Swingle16}; these are described by fixed points which are multicritical. An instance of a generic QPT that has $z\neq 1$, but which is weakly coupled and has a gapless phase next to it, is the Pokrovsky-Talapov transition \cite{pokrovsky1978phase, pokrovsky1979ground}, with $z=2$.

\begin{figure}[htb]
\includegraphics[width=\linewidth]{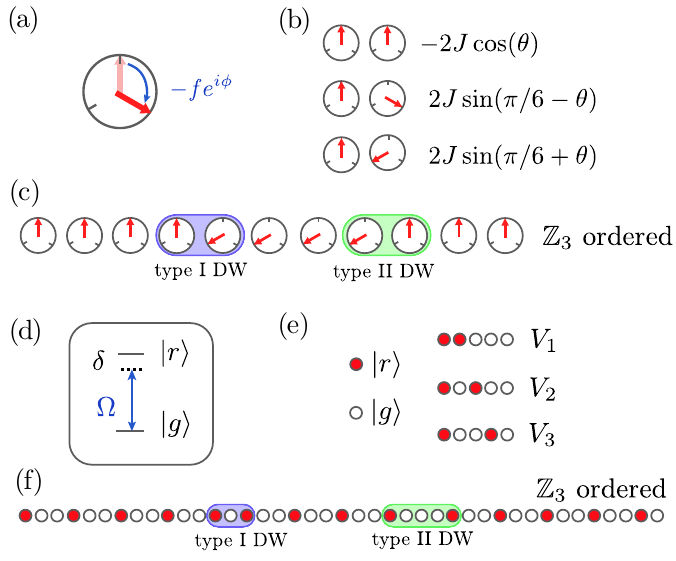}
\caption{\label{fig:Fig1}Schematic representation of (a--b) the interactions, and (c) a generic state of the $\mathbb{Z}_3$ chiral clock model. The arrows connote the eigenvalue of the operator $\sigma$ at each site with $\sigma_i = 1,\, \omega,\, \omega^2$ delineated by the arrow pointing at 12-, 4-, and 8-o'clock, respectively. Owing to the chirality of the couplings in Eq.~\eqref{eq:Hamiltonian}, there are two distinct types of domain walls (DW) with their associated interaction strengths illustrated in (b). (d) The Rydberg and ground states of the two-level system defined by Eq.~\eqref{eq:Rydberg}. The van der Waals interactions depend on the spacing between Rydberg excitations (e) and thus a representative state (f), once again, has two kinds of domain walls.}
\end{figure}

As implied above, generally, QPTs involving $\mathbb{Z}_n$ ($n \geq 3$) translational symmetry breaking along one spatial direction are  expected to be in the universality class of the $\mathbb{Z}_n$ {\it chiral} clock model \cite{huse1981simple,ostlund1981incommensurate,huse1982domain,huse1983melting,haldane1983phase,fendley2012parafermionic,zhuang2015phase}. The chirality is a consequence of the structure of domain walls between $\mathbb{Z}_n$-ordered phases: when moving along a fixed spatial direction, domain walls which move clockwise around the clock have distinct energies from those that move anti-clockwise [Fig.~\ref{fig:Fig1}]. There does not appear to be any proposed field-theoretic framework for understanding a generic direct transition from the gapped $\mathbb{Z}_n$-ordered phase to a gapped symmetric phase. The main impediment to such efforts is that the term in the action responsible for the chirality also induces incommensurate spin correlations, and in a perturbative analysis, the density-wave order parameter condenses at a nonzero wavevector, resulting in a state with long-range incommensurate order \cite{whitsitt2018}. Strong-coupling effects are evidently required to obtain a direct order--disorder transition, which is the focus of our study.

The $\mathbb{Z}_3$ chiral clock model (CCM) in one dimension is defined by the Hamiltonian \cite{ortiz2012dualities, fendley2012parafermionic}
\begin{equation}
\label{eq:Hamiltonian}
H_{\textsc{ccm}} = -f \,\sum_{j=1}^L \tau_j^\dagger \,\mathrm{e}^{-\mathrm{i}\,\phi} - J \sum_{j =1}^{L-1} \sigma_j^\dagger \,\sigma_{j+1}\,\mathrm{e}^{-\mathrm{i}\,\theta} + \mbox{h.c.}
\end{equation}
acting on a chain of $L$ spins; the Hilbert space is $(\mathbb{C}^3)^{\bigotimes L}$. The three-state spin operators $\tau_i$ and $\sigma_i$ act locally on the site $i$, and each satisfy 
\begin{align}
&\tau^3 = \sigma^3 = \II \,,
&\sigma\,\tau = \omega\,\tau\,\sigma\,; \quad \omega \equiv \exp\,(2\,\pi\,\mathrm{i}/3)\,.
\end{align}
For concreteness' sake, let us also explicitly choose the following representation of the CCM operators
\begin{equation}
\tau=
  \begin{pmatrix}
    1 & 0 & 0 \\
    0 & \omega & 0\\
    0 & 0 & \omega^2
  \end{pmatrix},
  \quad
\sigma=
  \begin{pmatrix}
    0 & 1 & 0 \\
    0 & 0 & 1\\
    1 & 0 & 0
  \end{pmatrix} \,,
\end{equation}
reminiscent of the Pauli matrices that measure and shift the spin at a given site. 
The scalar parameters $f$ and $J$ determine the on-site and nearest-neighbor couplings, while $\phi$ and $\theta$ define two chiral interaction phases. For $\phi$ and $\theta$ both nonzero, time-reversal and spatial-parity (inversion) symmetries are separately broken, but their product is preserved. This asymmetry in the Hamiltonian has important ramifications: the spatial chirality ($\theta\neq 0$) induces incommensurate (IC) floating phases with respect to the periodicity of the underlying lattice \cite{dai2017entanglement}. 
For applications to spatially ordered phases, we need $\phi=0$, whereupon time-reversal and spatial-parity are both symmetries of the Hamiltonian but the chirality is still present as a purely spatial one. This article is thus restricted to the $\phi=0$ case, with the chirality quantified by $\theta$ [Fig.~\ref{fig:Fig1}(b)].

The three-state CCM also has an explicit global $\mathbb{Z}_3$ symmetry. Using density-matrix renormalization group techniques, we study the critical behavior at the direct transition between the $\mathbb{Z}_3$-ordered and the gapped symmetric phase, with the aim of determining the exponent $z$. The achiral $\phi=0,\,\,\theta=0$ model has a transition in the universality class of the three-state Potts conformal field theory with $z=1$. We find that away from the special point $\theta=0$, the dynamical critical exponent $z$ is larger than 1, indicating that there is no emergent conformal invariance. For $\phi=0,\,\, \theta\neq 0$, our results [see Fig.~\ref{fig:iDMRG}] show that the gapped symmetric phase has spatially incommensurate correlations of the $\mathbb{Z}_3$ order parameter. However, the incommensurability vanishes as the transition is approached and the long-range $\mathbb{Z}_3$ order is eventually commensurate. These results clarify how a direct transition is possible between the gapped symmetric phase and $\mathbb{Z}_3$ order, without an intermediate gapless IC phase. 

Turning to the recent experiments with trapped Rydberg atoms \cite{bernien2017probing, labuhn2016tunable}, we consider a model which is directly related to the microscopic physical realization but the transitions of which are expected to be in the same universality class as in the corresponding $\mathbb{Z}_n$ CCM. On a microscopic level, a one-dimensional array of $N$ atoms is described by the Hamiltonian 
\begin{align}\label{eq:Rydberg}
H_{\rm Ryd}=&\sum_{i=1}^N \frac{\Omega}{2} (\ket{g}_i\!\bra{r}+\ket{r}_i\!\bra{g})-\delta \ket{r}_i\!\bra{r}\nonumber\\
&+\sum_{i<j} V_{|i-j|}\ket{r}_i\!\bra{r}\otimes \ket{r}_j\!\bra{r}.
\end{align}
Here, $\ket{g}_i$ and $\ket{r}_i$ denote the internal atomic ground state and a highly excited Rydberg state of the $i^\mathrm{th}$ atom, which together represent a spin-1/2 system [Fig.~\ref{fig:Fig1}(d--f)]. The parameters $\Omega$ and $\delta$ characterize a coherent laser driving field, while $V_{x} = C_6/x^6$ quantifies the van der Waals interactions of atoms in Rydberg states. In this study, we focus on a region in parameter space where this system exhibits a QPT between the $\mathbb{Z}_3$-ordered and the gapped symmetric phase \cite{fendley2012parafermionic} and provide numerical evidence that the critical behavior parallels that of the three-state CCM \eqref{eq:Hamiltonian}. We note that $H_{\rm Ryd}$ does not break time-reversal symmetry, motivating our choice of  $\phi=0$ in the study of the quantum clock model \eqref{eq:Hamiltonian}.

Beginning with the chiral clock model in Sec.~\ref{sec:model}, we outline the approach used to determine the critical exponents of phase transitions for both $H_\textsc{ccm}$ and $H_{\rm Ryd}$ in Sec.~\ref{sec:formalism}.
Numerical results then follow: for the CCM in Sec.~\ref{sec:model}, and the Rydberg model (which can be mapped to a system of hard-core bosons \cite{sachdev2002mott, fendley2004competing, GSS18}) in Sec.~\ref{sec:boson}. The associated exponents, which we compile in Secs.~\ref{sec:P0} and \ref{sec:exp_UV}, respectively, are shown to differ rather nontrivially from the Potts exponents and underscore the nonconformal nature of the critical field theory. We also appraise the possible influence of long-range interactions on the nature of the chiral $\mathbb{Z}_3$ transition in Sec.~\ref{sec:lr}. Finally, we conclude by summarizing our results in Sec.~\ref{sec:end}.

\section{The $\mathbb{Z}_3$ chiral clock model}
\label{sec:model}

The one-dimensional $\mathbb{Z}_n$ clock model \cite{ostlund1981incommensurate, huse1981simple, huse1982domain, huse1983melting} is a straightforward generalization of the transverse-field Ising model (TFIM) wherein one replaces the two-state Ising spin at each site by a variable with $n$ states. However, instead of enlarging the symmetry from $\mathbb{Z}_2$ to that of the permutation group $\mathbb{S}_n$ (which would result in the $n$-state Potts model), one can construct a Hamiltonian such that the interactions are invariant under only the subgroup $\mathbb{Z}_n$. Confining ourselves to $n=3$, this leads to the $\mathbb{Z}_3$ clock model \cite{fendley2012parafermionic} in Eq.~(\ref{eq:Hamiltonian}), where the three values of the spin at any site are most conveniently labeled by $1,\,\omega,\,\omega^2$. In this representation, it is not difficult to observe that the Hamiltonian \eqref{eq:Hamiltonian} is invariant under a uniform rotation of all the spins as $\tau_j \rightarrow \tau_j$, $\sigma_j \rightarrow \omega\,\sigma_j \,\,\forall\, \,j$ and thereby possesses a global $\mathbb{Z}_3$ symmetry implemented by $\mathcal{U} = \prod_{j=1}^L \tau_j^{\dagger}$. 

This model's recent resurgence, some forty years after its original proposal, is closely tied to the presence of non-Abelian bound states beyond Majorana fermions, after a non-local mapping of the Hilbert space to those of parafermions (described below). 
Both analytical \cite{fendley2012parafermionic} and numerical \cite{jermyn2014stability} studies confirm this assertion: parafermionic edge zero modes can and do exist in this model \cite{moran2017parafermionic} but only when the interactions are chiral; in contrast, the more conventionally studied clock models with purely ferromagnetic or antiferromagnetic interactions do not boast boundary zero modes.

In analogy with the TFIM, the Hamiltonian of the 1D CCM in Eq.~(\ref{eq:Hamiltonian}) comprises a spin-flip term ($f>0$) and a two-site interaction term ($J>0$). By continuously varying the phases $\phi$ and $\theta$, the effective interaction can be interpolated between ferromagnetic (e.g., $f=J=0.5$, and $\phi = \theta = 0$) and antiferromagnetic ($\phi = \theta = \pi/3$). The CCM is known to be integrable \cite{au1997many, baxter2006challenge} for a two-parameter family of couplings \cite{au1987commuting} along the line $f \,\cos \,(3\, \phi) = J \,\cos \,(3 \, \theta)$ and at precisely $\phi = \theta = \pi/6$, it is superintegrable \cite{albertini1989commensurate, mccoy1990excitation}. On setting $\phi = \theta = 0$, it reduces to just the quantum version of the three-state Potts model.  

\subsection{Parafermionic description}

Recalling the reformulation of the TFIM in terms of Majorana fermions, it is natural to wonder if a similar procedure could be carried out for the $\mathbb{Z}_3$ CCM too. As illustrated by \citet{fendley2012parafermionic}, the answer is indeed in the affirmative: the Fradkin-Kadanoff transformation \cite{fradkin1980disorder} maps the model onto a parafermionic chain 
\begin{equation}
H_3 = -f \,\sum_{j=1}^L \psi_j^\dagger \,\chi_j\,\mathrm{e}^{-\mathrm{i}\,\phi} - J\,\omega^2 \sum_{j =1}^{L-1} \psi_j^\dagger \,\chi_{j+1}\,\mathrm{e}^{-\mathrm{i}\,\theta} + \mbox{h.c.}
\end{equation}
where the two basic parafermions are 
\begin{equation}
\label{eq:pf}
\chi_j = \left ( \prod_{k=1}^{j-1} \tau_k \right) \sigma_j, \quad \psi_j = \omega \,\left ( \prod_{k=1}^{j-1} \tau_k \right) \sigma_j \,\tau_j.
\end{equation}
These operators neither square to one nor commute:
\begin{equation*}
\chi_j^3 = \tau_j^3 = \II;\,\,\, \chi_j^\dagger = \chi_j^2;\,\,\, \psi_j^\dagger = \psi_j^2;\, \mbox{ and } \, \chi_j\,\psi_j = \omega\, \psi_j \chi_j.
\end{equation*}
Additionally, due to to the inherently nonlocal nature of the duality transformation \eqref{eq:pf}, the parafermionic operators at different points also do not commute but instead satisfy the algebra
\begin{equation}
\chi_j \chi_k = \omega\, \chi_k\chi_j;\,\,\, \psi_j \psi_k = \omega\, \psi_k\psi_j;\,\,\, \chi_j\,\psi_k = \omega\, \psi_k \chi_j,
\end{equation}
for $j < k$. The most striking feature that manifests itself upon this redefinition is the occurrence of edge zero modes characteristic of topological order \cite{fendley2012parafermionic, fendley2014free}. These zero-energy modes, which are robust even with disordered couplings (i.e., spatially varying $f$ and $J$), require the interactions to be necessarily chiral, unlike in the usual ferromagnetic and antiferromagnetic cases.

\subsection{Phase diagram}

The exact phase structure of the quantum clock model has been a subject of much debate. Intuitively, one expects to find a disordered ($\mathbb{Z}_3$-symmetric) and an ordered (symmetry-broken) phase---as is the case with the TFIM---with a transition between the two driven by tuning the coupling strengths from a regime where $f \gg J$ to one where $f \ll J$. Studies on several variants of this model, using a multitude of methods spanning from Monte Carlo simulations to transfer matrix partial diagonalization \cite{selke1988annni, von1984finite, everts1989transfer}, lent credence to this expectation. Early analyses of finite-size effects \cite{von1984finite} (admittedly, restricted to $\theta = 0$) revealed an additional IC phase absent in the TFIM with chiral interactions \cite{ostlund1981incommensurate}. Incidentally, although it was known that the model has a tricritical Lifshitz point \cite{straley1973three, hornreich1975critical} at the intersection of the three phases, its precise location in parameter space was, until quite recently, ambiguous \cite{selke1982monte, howes1983commensurate}.

\begin{figure}[htb]
\includegraphics[height=6cm]{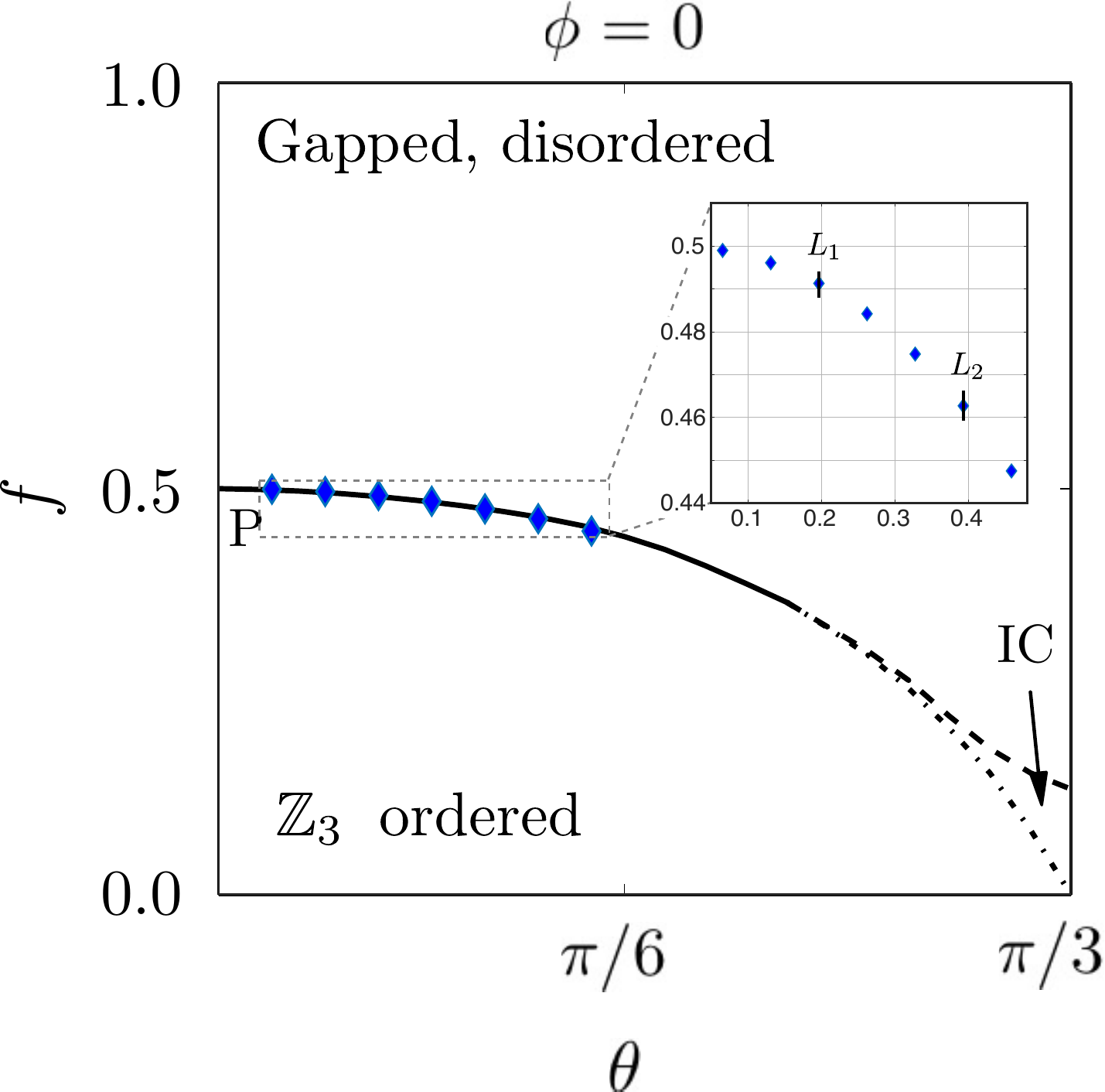}
\caption{\label{fig:CCM_PD}The three phases of the CCM, adapted from Ref.~\onlinecite{zhuang2015phase}, are schematically limned in a cross-section of the three-dimensional phase diagram for $\phi = 0$, $J = 1-f$, and $L = 100$. $P$ marks the transition point of the three-state Potts model. Blue diamonds indicate the quantum critical points obtained from finite-size DMRG, which are listed in Table~\ref{Table:z}. When overlaid with the previously found transition points (solid line), the two datasets agree perfectly.}
\end{figure}

The phase diagram of the $\mathbb{Z}_3$ CCM was mapped out as a function of both chiral parameters ($\phi$, $\theta$) and couplings ($f$, $J$) by \citet{zhuang2015phase}, who detected three distinct phases from entanglement entropy (EE) considerations \cite{calabrese2004entanglement, pollmann2010entanglement}. We show the phase diagram for the case with $\phi=0$ in Fig.~\ref{fig:CCM_PD}. For large (small) values of $f$, the system is generically in the disordered (ordered) state. In the parafermionic description, these can be designated trivial and topological, respectively, the latter being characterized by a threefold ground-state degeneracy (ergo, an EE $\sim \ln\, 3$). Both these phases are gapped and are separated by an intermediate IC region when $\theta$ is large $\left(\gtrsim \pi/6 \right)$. The disorder--IC transition is of the Kosterlitz-Thouless type \cite{kosterlitz1973ordering} while the order--IC one is a Pokrovsky-Talapov transition \cite{pokrovsky1978phase, pokrovsky1979ground}. For small $\theta$, our results in Section.~\ref{sec:P0} support a 
direct phase transition between the two gapped phases (shown in Fig.~\ref{fig:CCM_PD}), as proposed by earlier works \cite{selke1982monte, howes1983quantum,howes1983commensurate,zhuang2015phase}. Arguments that a gapless IC phase must intervene between the gapped phases \cite{haldane1983phase} do not apply near the Potts point at $\theta=0$.
Little is known about the universality class of the direct $\mathbb{Z}_3$-symmetry-breaking (or in parafermionic language, direct trivial--topological) transition and hence, understanding it, in its full generality, is our desideratum in this section. What we do know, however, is that the clock Hamiltonian \eqref{eq:Hamiltonian} exhibits a second-order QPT at $f = J$ when $\theta = \phi = 0$. The symmetry group for the (achiral) model at this critical point is the full $\mathbb{S}_3$ and the underlying critical conformal field theory is that of the three-state Potts model with central charge 4/5 \cite{ginsparg1989fields}. 
%Since the system is self-dual for $f=J$ and $\phi = \theta$, the boundary for a trivial--topological phase transition, if it exists, must necessarily lie on this line. 

Our general strategy is therefore as follows. We sweep across the phase transition, constrained to the subspace $f = 1-J$, at several discrete values of $\theta$ for $\phi = 0$. At each such point in parameter space, the ground-state eigenvalue $E_0$ and the energy gap to the nearest level $\Delta = E_1 - E_0$ are recorded. Since the form of the Hamiltonian is invariant under either of the transformations
\begin{alignat}{2}
\phi' &\rightarrow \phi + \frac{2\,m\,\pi}{3}; \quad &&\theta' \rightarrow \theta + \frac{2\,n\,\pi}{3};\,\, \forall\,\, m, n \in \mathbb{Z},\\
\phi' &\rightarrow - \phi; \quad &&\theta' \rightarrow -\theta,
\end{alignat}
it suffices to consider only the range $\phi, \,\theta \le \pi/3$ and in particular, $\phi, \,\theta < \pi/6$, where a \textit{direct} transition has been confirmed \cite{zhuang2015phase}. Systematically examining this region enables us to find the critical exponents of the transition as detailed below. 

\subsection{Critical exponents}
\label{sec:formalism}
Of primary interest in characterizing the nature of the $\mathbb{Z}_3$-symmetry-breaking phase transition are its critical exponents. Specifically, these include the dynamical critical exponent $z$ and the correlation length exponent $\nu$, which are defined by \cite{sachdev2011quantum, dutta2015quantum}
\begin{alignat}{1}
\label{eq:def}
\Delta &\sim \left \lvert f - f_c \right \rvert^{z\,\nu};\qquad
\xi \sim \left \lvert f - f_c \right \rvert^{-\,\nu}, 
\end{alignat}
where $\Delta$ denotes the mass gap and $\xi$ is the correlation length that governs the decay of $G (r) \equiv \langle \Psi (r) \, \Psi(0) \rangle -\langle \Psi (r)  \rangle \langle \Psi (0) \rangle \sim \exp(-r/\xi)$ for an order parameter $\Psi$. In order to numerically estimate these exponents, we use finite-size scaling (FSS) \cite{fisher1972scaling, hamer1980finite} as described below.

\subsubsection{Finite-size scaling and extrapolation}
\label{sec:FSS}
The FSS approach posits that if some thermodynamic quantity $\mathcal{K}\, (f)$ diverges in the bulk system as $\mathcal{K} \,(f) \sim \lvert f - f_c \rvert^{-\kappa}$ as $f \rightarrow f_c$, then, at criticality, it scales as $\mathcal{K}\, (f_c) \sim L^{\kappa /\nu}$ on a lattice of $L$ sites. The exponent $\kappa/\nu$ can therefore be directly determined by plotting $\mathcal{K}$ against the system size. The FSS procedure now hinges on appropriate choices of the variable $\mathcal{K}$. For instance, near the quantum critical point (QCP), one can assume that the gap obeys a scaling ansatz of the functional form
\begin{equation}
\label{eq:ansatz}
\Delta = L^{-z} \, \mathcal{F} \left( L^{1/\nu}\,(f-f_c) \right),
\end{equation}
with $\mathcal{F}$ some universal scaling function. In addition, from a technical perspective, it is also helpful to define the Callan-Symanzik $\beta$ function \cite{hamer1979strong}:
\begin{equation}
\label{eq:beta}
\beta\, (f) = \dfrac{\Delta\, (f)}{\Delta\, (f)  - 2\, {\displaystyle \dfrac{\partial\,\Delta (f) \vphantom{\huge e^{x^x}}} {\partial\, \ln f}} }.
\end{equation}
From Eqs.~\eqref{eq:ansatz} and \eqref{eq:beta}, it follows that these two quantities scale as $-z$ and $-1/\nu$, respectively, at the QCP thus giving us access to the required exponents. 

Alternatively, starting from successive values of $\mathcal{K}$ on a sequence of finite lattices of increasing size, one can estimate the ratio $\kappa/\nu$ from the sequence
\begin{equation}
\label{eq:Ratio}
R_{\mathcal{K}, L} = \frac{L\,\left[\mathcal{K}_L (f_c) - \mathcal{K}_{L-2}(f_c) \right]}{2\,\mathcal{K}_{L-2}(f_c)},
\end{equation}
which converges to $\kappa/\nu$ as $L \rightarrow \infty$ \textit{if} there are no other higher-order or offset corrections. Under this assumption, Eq.~\eqref{eq:Ratio} asserts that in the limit $L \rightarrow \infty$,  $R_{\Delta, L} \rightarrow -z$. On a finite lattice, however, this limit is approached only asymptotically, necessitating further extrapolation techniques \cite{hamer1981finite} to accelerate the convergence of the sequences \eqref{eq:Ratio} and  to improve the estimates of the critical parameters. In this work, wherever applicable, we employ the BST \cite{bulirsch1964, henkel1988finite} algorithm to this end. 

\subsubsection{Numerical results}
\label{sec:P0}

Our numerical calculations in this section are based on the density-matrix renormalization group (DMRG) algorithm \cite{white1992density, white1993density,ostlund1995thermodynamic, rommer1997class, dukelsky1998equivalence,peschel1999density}. We use finite-system DMRG \cite{schollwock2005density, schollwock2011density} with a bond dimension $m = 150$ for a chain of up to $L=100$ three-state spins with open boundary conditions; the first and second energy levels are individually targeted. The energy eigenvalues were found to be reasonably converged within three sweeps to an accuracy of one part in $10^{10}$. 

First of all, let us inspect the QPT at the Potts point, $\phi = \theta = 0$ and $f_c = 0.5$. Using Eq.~\eqref{eq:Ratio} recursively, we compute $R_{\Delta, 100} = 0.9881$, whereupon extrapolation to its infinite-lattice value yields $z = 0.9959$. Next, we consider the scaling of the Callan-Symanzik $\beta$ function [Eq.~\ref{eq:beta}, Fig.~\ref{fig:Beta}] in connection with extracting the correlation length exponent. Evaluating the slope of a log-log plot against $L$, we find that for $\phi = \theta = 0 $, 
$\nu^{-1} = 1.201\pm 0.001$ or, in other words, $\nu = 5/6$ as is known independently for the three-state Potts model. It is to be noted that both values are in excellent agreement with the universality hypothesis \cite{alexander1975lattice} that the three-state Potts model and Baxter's ``hard-hexagon'' model \cite{baxter1980hard} should have the same critical exponents. Reassured by these familiar exponents, we now look into the critical parameters for nonzero chiral angles. 
%It is useful to note that both the Hamiltonians  $H\, (\phi,\, \theta = 0)$ and $H\, (\phi = 0,\, \theta)$ have separate time-reversal and parity symmetries \cite{mong2014parafermionic}, though their explicit definitions in the two regimes may differ. In contrast, when both $\phi$ and $\theta$ are nonzero, the only discrete symmetry is CPT. 

\begin{figure}[htb]
\subfigure[]{\label{fig:a}\includegraphics[width=\linewidth]{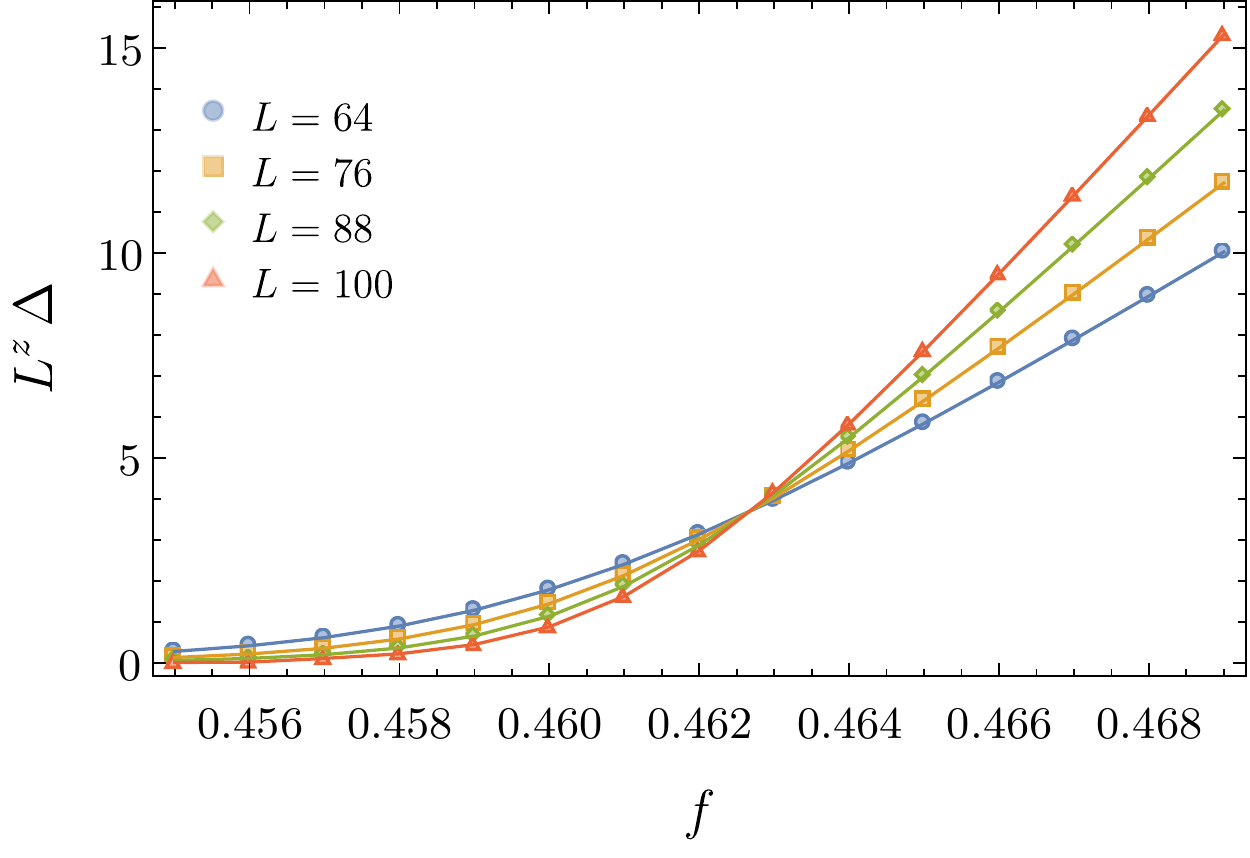}}
\subfigure[]{\label{fig:b}\includegraphics[width=0.4775\linewidth]{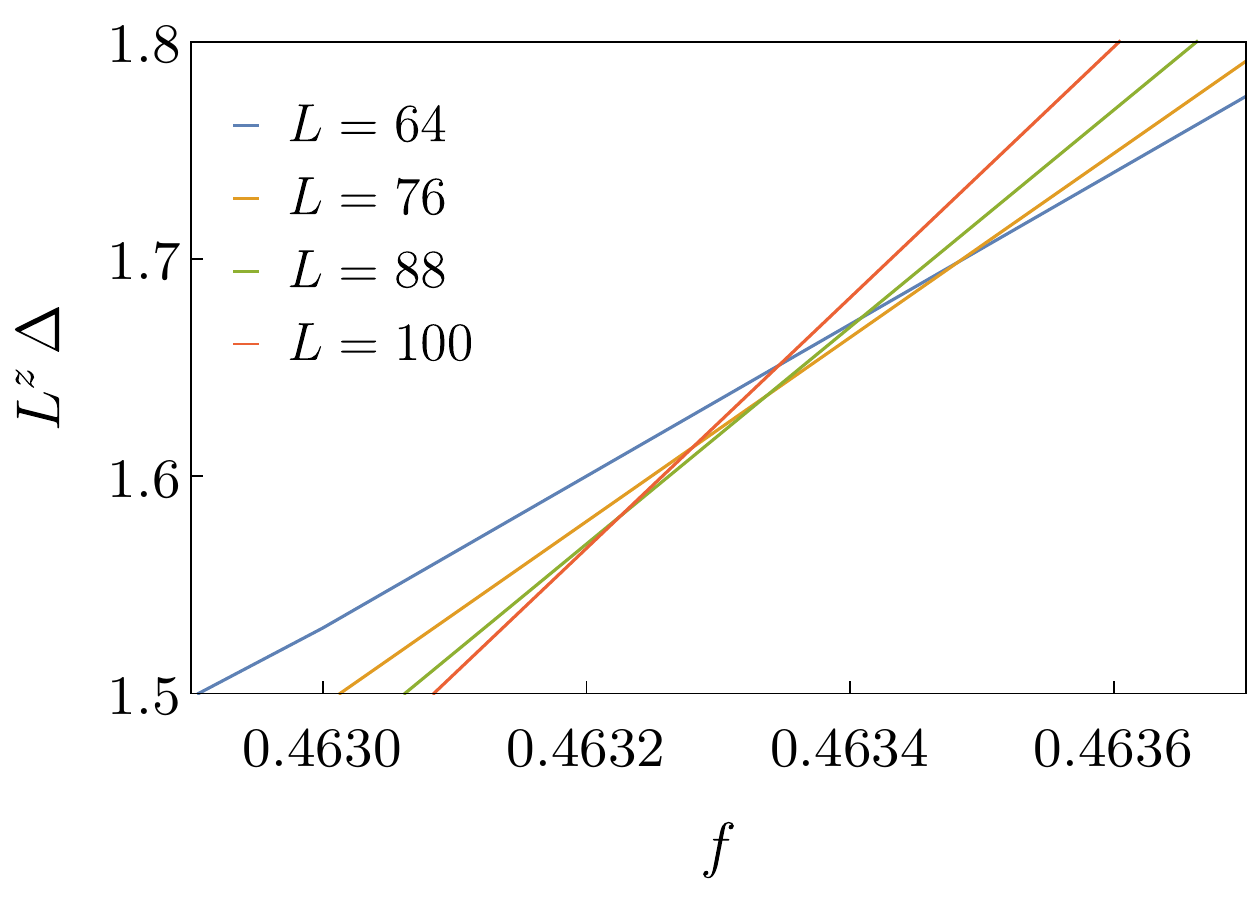}}
\subfigure[]{\label{fig:c}\includegraphics[width=0.5025\linewidth]{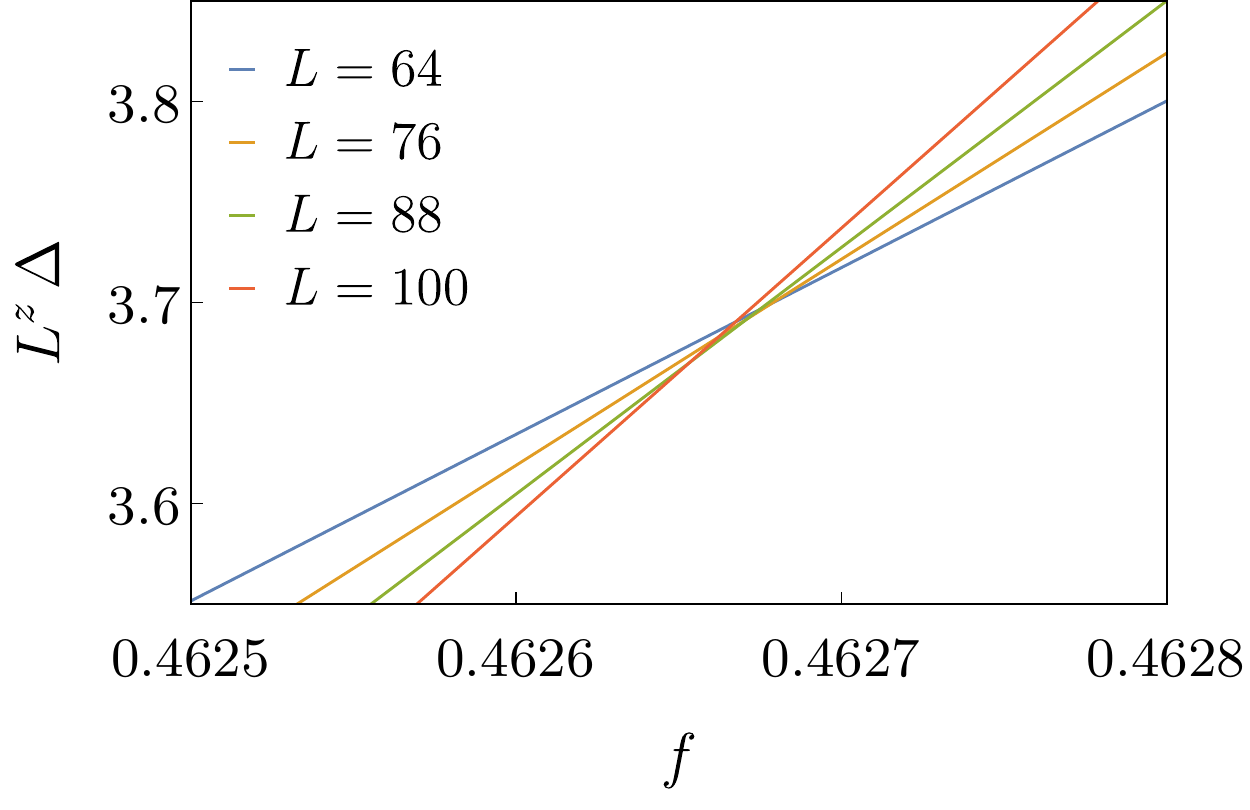}}
\caption{\label{fig:Cross}(a) Scaling of the energy gap $\Delta$ as a function of $f$ for individual system sizes with $\phi = 0$ and $\theta = \pi/8$. With $z = 1.229$, all the curves intersect right at the critical point. The finesse of the crossing depends crucially on the $z$ chosen: on zooming in, the contrast in sharpness between $z=1$ (b) and $z = 1.229$ (c) is vivid. }
\end{figure}

In the remainder of this section, we focus on the $\mathbb{Z}_3$-symmetry-breaking transition for a broad range of values of $\theta$, which we adjust here between $0 \le \theta < \pi/6$ in steps of $\pi/48$. Constrained by the lack of preexisting information about $z$, we pinpoint the location of the QPT by plotting $L^z \, \Delta_L$ against the tuning parameter $f$ for various lattice sizes (ranging from $L=60$ to $L=100$) and values of $z$ [see Fig.~\ref{fig:a}]. Precisely at the QCP $f=f_c$, Eq.~\eqref{eq:ansatz} predicts that the quantity $L^z\Delta$ is independent of the length of the system $L$. In other words, with the correct choice of $z$, all the curves for $L^z\Delta$ should cross at $f_c$ for different values of $L$. This allows us to both locate the QCP $f_c$ \textit{and} estimate the value of $z$ simultaneously. By scanning successively smaller intervals, we are able to determine the intersection point of the curves for different lengths to an accuracy of $10^{-4}$. Fig.~\ref{fig:Cross} displays an example of this method for $\theta = \pi/8$. The variation of the crossing points with $\theta$ is noted in Table~\ref{Table:z}; the positions of the QPT calculated in this fashion are in exact agreement with the phase boundaries reported by Ref.~\onlinecite{zhuang2015phase}.

\begin{figure}[htb]
\includegraphics[width=\linewidth]{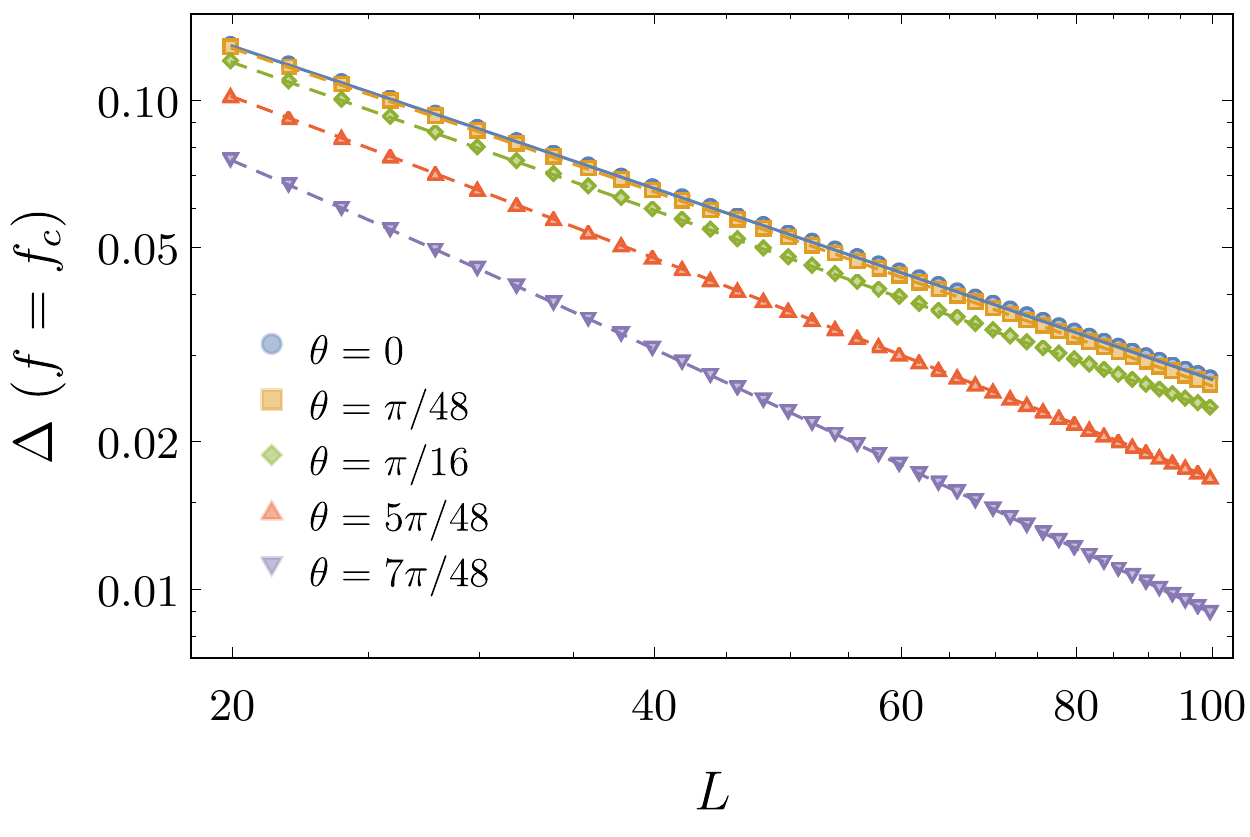}
\caption{\label{fig:FSSP=0}Log-log plot of the gap $\Delta$ against the lattice size $L$ at the critical point $f=f_c$. As previously, $\phi = 0$, and the different values of $\theta$ are labeled. For sufficiently large systems, the dependence is exactly linear as expected. For reference, the curve for the Potts transition is indicated by a solid line. }
\end{figure}
\begin{figure}[htb]
\includegraphics[width=\linewidth]{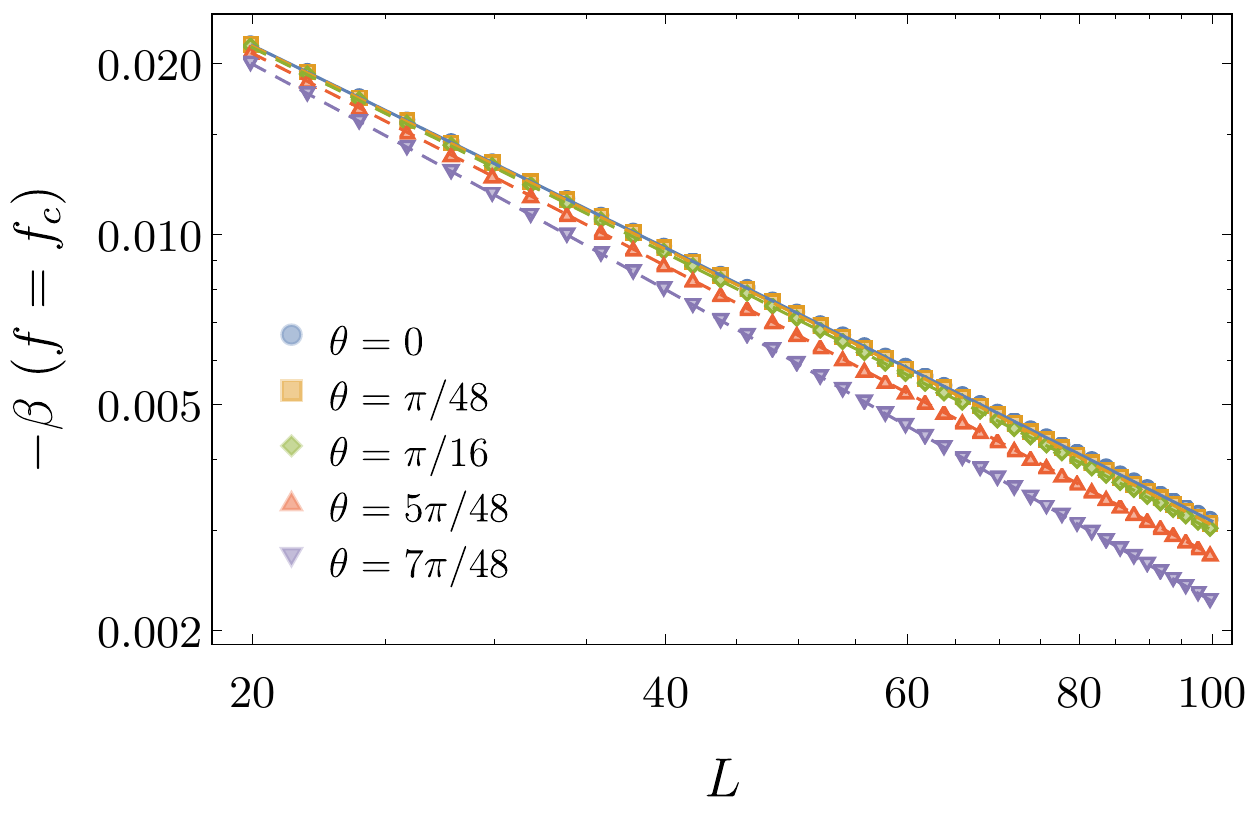}
\caption{\label{fig:Beta}The Callan-Symanzik $\beta$ function plotted on a logarithmic scale against the system size for $\phi = 0$. The slopes of the curves in the linear region, corresponding to large lattices, give us the respective values of $-1/\nu$ for each choice of $\theta$.}
\end{figure}
Although the abovementioned approach kills two birds with one stone, one could worry about sensitivity to the range of system lengths over which FSS is applied. To carefully investigate any such effects, we now turn our attention to the scaling of the mass gap at precisely $f=f_c$: the gap is presented as a function of the system size in Fig.~\ref{fig:FSSP=0}. The deviations from a pure power-law scaling bespeak the presence of corrections that become important for small spin chains. In order to incorporate corrections to FSS for the datasets, we use the ansatz 
\begin{equation}
\label{eq:DelCorr}
\Delta \,(L) = c_0\, L^{-z} (1 + c_1\, L^{-\zeta} ); \quad (\zeta > z),
\end{equation}
that was originally argued for the three-state quantum Potts chain on the premise of conformal invariance \cite{von1987conformal, reinicke1987analytical}. The coefficients $c_i$ as well as the exponents $z$ and $\zeta$ are treated as free parameters for the fit. The values of $z$ thus obtained are enlisted in Table~\ref{Table:z}, which shows that $z>1$ for $\theta > 0$, implying that the underlying field theory breaks Lorentz invariance and (as hinted at in Sec.~\ref{sec:boson} as well) is \textit{not} conformally invariant. Hence, for $\theta \ne 0$, the QPT is not in the same universality class as the ordinary order--disorder transition in the  three-state Potts model. Earlier examples of strongly coupled quantum critical theories with $z \neq 1$ are limited to quantum critical metals \cite{lee2017recent} and the onset of Ising ferromagnetism in one-dimensional metals \cite{SS96,KY04,Sau17}. Interestingly, even for the \textit{classical} asymmetric clock model in two dimensions, spatial rotational symmetry is not recovered in the scaling limit (and consequently, it cannot be described by a conformal field theory) \cite{henkel2013conformal}.

\bgroup
\def\arraystretch{1.75}
\begin{table}[tb]
%\vspace*{0.4cm}
\centering
{
\bgroup
\setlength{\tabcolsep}{11.25pt}
\begin{ruledtabular}
\begin{tabular}{l l l l l} 
\multicolumn{1}{c}{$\theta$} &\multicolumn{1}{c}{$f_c$} &\multicolumn{1}{c}{\large $z$}  &\multicolumn{1}{c}{$\mbox{\large $z$}_{\,\zeta}$} &\multicolumn{1}{c}{$1/\nu$}  \\ \hline
$\pi/48$  & $0.4990$ & $1.003$ &$1.00(7)$ & $1.20(9)$\\
$\pi/24$  & $0.4961$ & $1.021$ &$1.01(8)$ & $1.21(8)$  \\
$\pi/16$  & $0.4913$ & $1.022$ &$1.02(1)$ & $1.22(3)$ \\
$\pi/12$  & $0.4842$ & $1.078$ &$1.07(6)$ & $1.25(1)$ \\
$5 \pi/48$  & $0.4748$ & $1.135$ &$1.13(3)$ & $1.27(7)$ \\
$\pi/8$  & $0.4627$ & $1.229$ & $1.22(7)$ & $1.32(4)$ \\
$7\pi/48$ &$0.4475$ & $1.368$ &$1.36(6)$ & $1.38(2)$ \\
\end{tabular}
\end{ruledtabular}
}
\egroup
\caption{\label{Table:z}Calculated dynamical and correlation length critical exponents for $\phi = 0$. Two independent sets of values of $z$ are distinguished: the first series is our estimate from the crossing points [Fig.~\ref{fig:Cross}] whereas the second (designated by the subscript $\zeta$) is for the values after correcting for finite-size effects, determined from a nonlinear fit to Eq.~\eqref{eq:DelCorr}. The evolution of the exponents with $\theta$ is quite smooth.}
\end{table}
\egroup

\begin{figure*}[htb]
\includegraphics[width=0.33\linewidth]{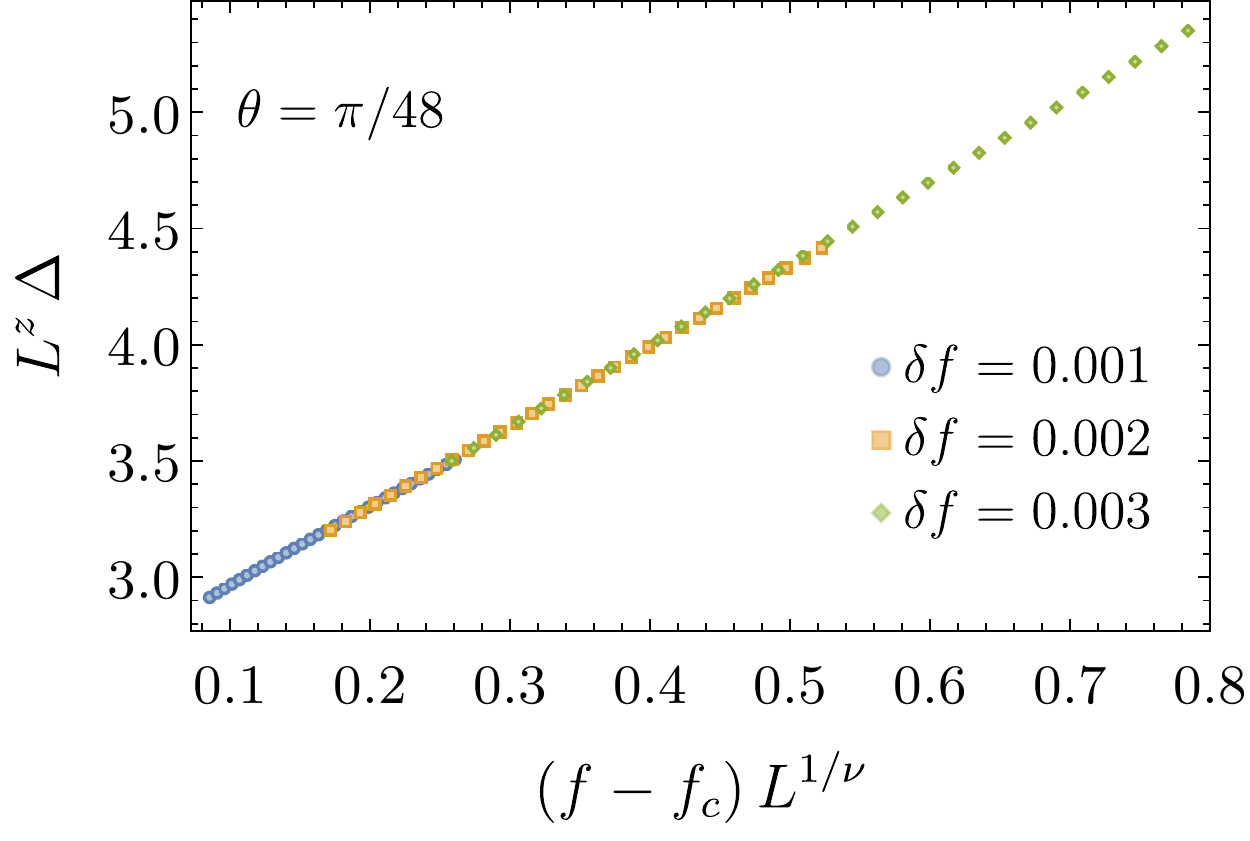}
\includegraphics[width=0.325\linewidth]{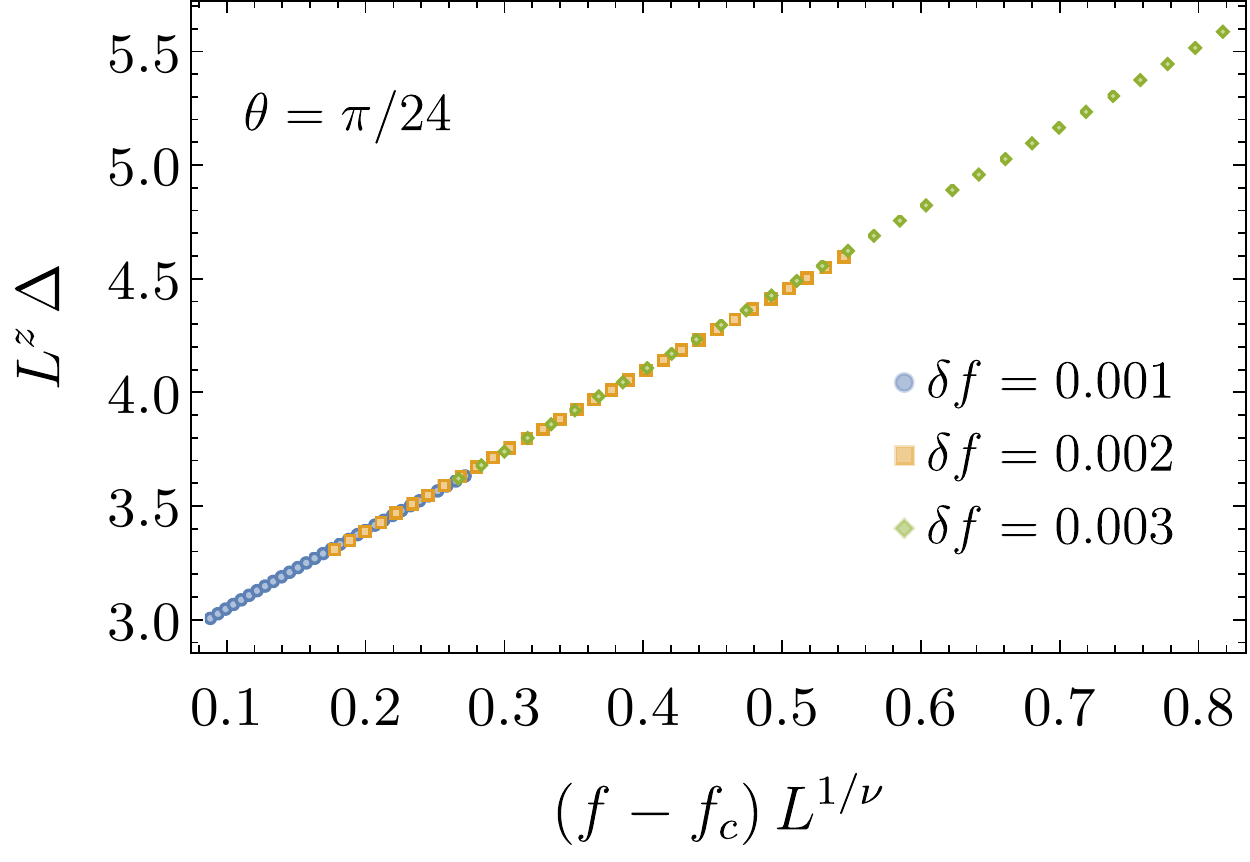}
\includegraphics[width=0.325\linewidth]{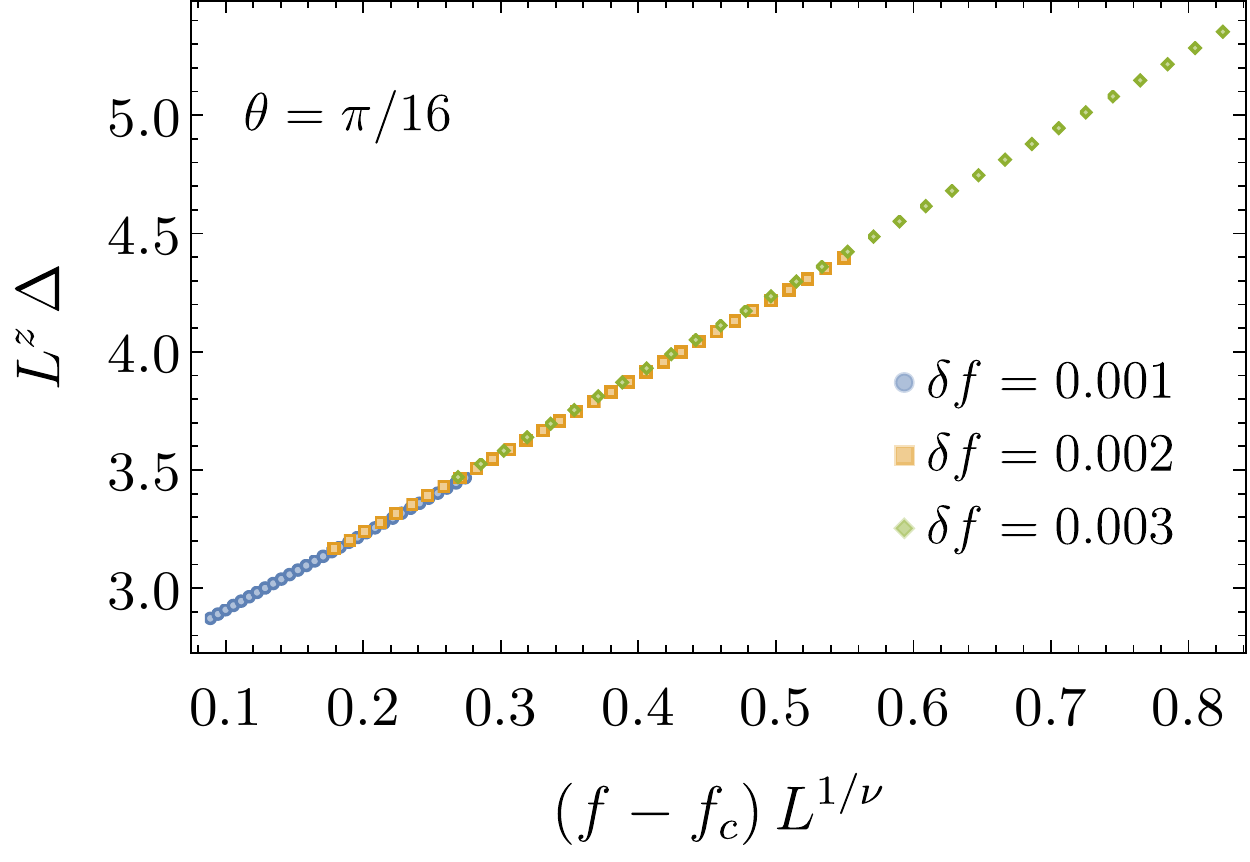}
\includegraphics[width=0.33\linewidth]{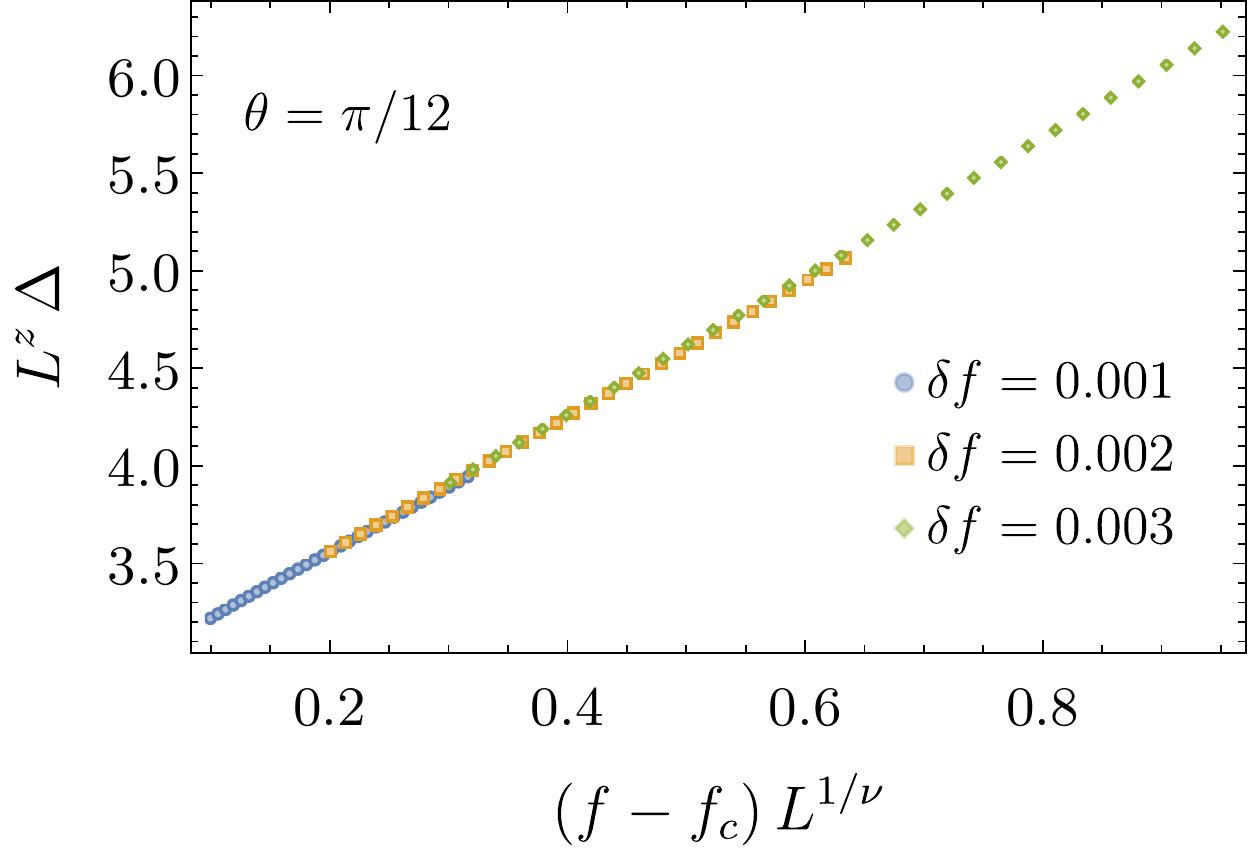}
\includegraphics[width=0.325\linewidth]{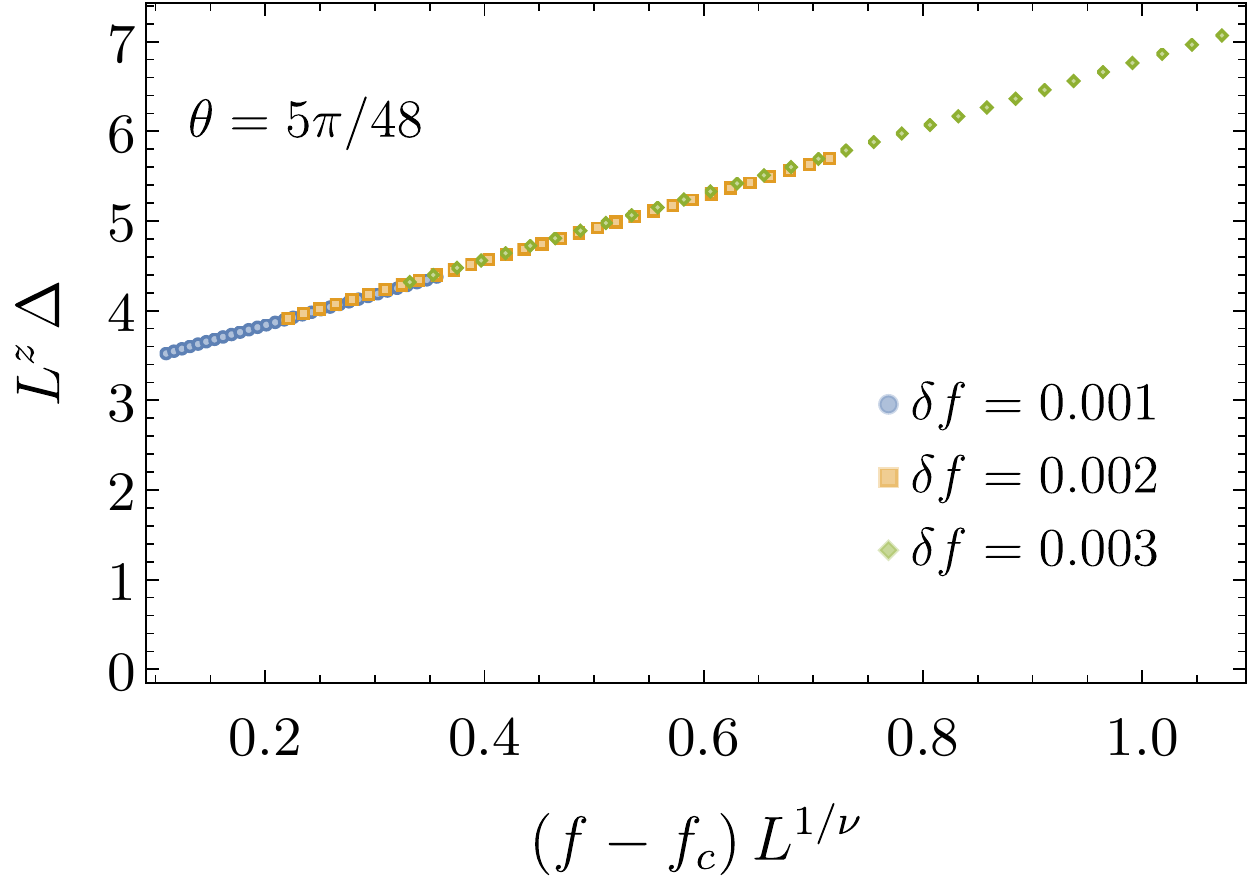}
\includegraphics[width=0.325\linewidth]{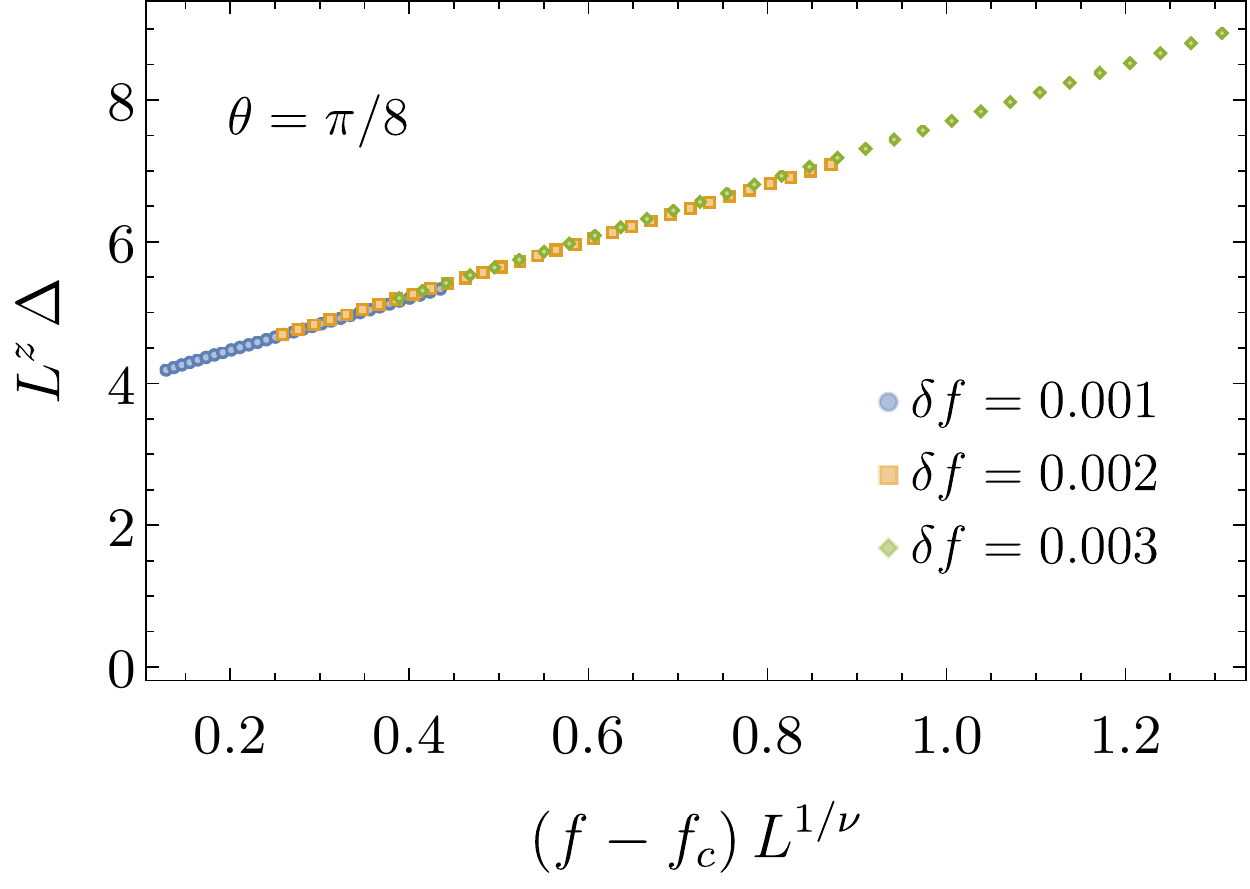}
\caption{\label{fig:Collapse}Data collapse for six different chiral angles based on the critical exponents $z$ and $\nu$ recorded in Table~\ref{Table:z}. The data consists of three different sets, one for each value of $\delta f \equiv f - f_c$. Every dataset is, in turn, comprised of 31 points for the differing system lengths $L$ ranging from 40 to 100, in steps of 2. All the data collapse perfectly onto a single line.}
\end{figure*}

Table~\ref{Table:z} also lists the correlation length exponents observed for different angles. It is perhaps not altogether surprising \cite{centen1982non} that $\nu$ varies continuously with the parameter $\theta$; in this case, it ranges between $5/6$ and (roughly) $5/7$, which is in consonance with the exponents of the Rydberg array model studied in the following section. The robustness of all these exponents can be independently verified by yet another method. Eq.~\eqref{eq:ansatz} stipulates that on graphing $L^z\,\Delta$ as a function of the combined scaling variable $(f-f_c)\, L^{1/\nu}$, all the data for different values of $f$ and $L$ should collapse onto a single curve \cite{landau1976finite, binder1980phase, bhattacharjee2001measure}. Using the values of $z$ and $\nu$ from Table~\ref{Table:z}, we obtain excellent data collapse as established by Fig.~\ref{fig:Collapse}.

Altogether, the analysis presented here offers strong evidence for a direct second-order transition from the disordered to the commensurate $\mathbb{Z}_3$-ordered phase.
Then, a natural question as regards scaling is whether the direct transition is described by a single fixed point with a universal set of critical exponents and scaling functions, or whether these exponents vary continuously.
The overwhelming majority of second-order transitions are known to be of the former kind, with fixed exponents. All familiar examples of the latter are related to the Kosterlitz-Thouless transition and unfortunately, based on our data, this category cannot be conclusively ruled out. On the contrary, the variation in the values of $z$ and $\nu$ could also be an artifact stemming from finite-size effects.

\begin{figure*}[htb]
\begin{center}
\includegraphics[width = \linewidth]{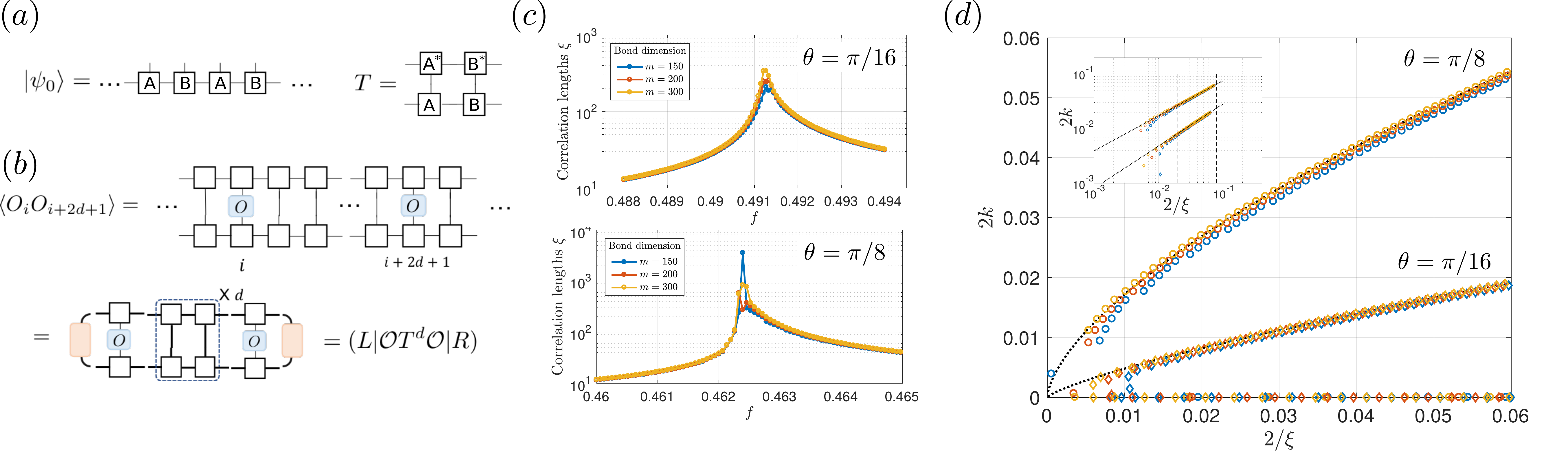}
\caption{\label{fig:iDMRG}Numerical simulations using iDMRG. Diagrammatic representations of (a) two-site-translation invariant matrix product states and two-site transfer matrix $T$, and (b) the two-point correlation function. (c) Correlation lengths $\xi$ along the two lines $L_1$ ($\theta = \pi/16$) and $L_2$ ($\theta = \pi/8$,) indicated in the inset of Fig.~\ref{fig:CCM_PD}, for three different bond dimensions. The divergence of the correlation lengths signal the onset of the QPT.
(d) The wavevector $k$, defined in Eq.~\eqref{eq:T_eig}, tends to zero as $\xi \rightarrow \infty$. However, the scaling form upon approaching the critical point is different for different chiral angles. Note that the points at the bottom, along the $x$-axis, correspond to values of $\lambda_1$ \textit{inside} the symmetry-broken phase;
the concomitant momenta are exactly zero.}
\end{center}
\end{figure*}

In order to obtain some further insight into the nature of this phase transition, we probe the correlation properties of the ground states near the critical point by performing infinite-system DMRG (iDMRG) \cite{mcculloch2007density, mcculloch2008infinite} calculations. More specifically, we sequentially optimize the ground-state wavefunction assuming an infinite-size matrix product state (iMPS) that is invariant under two-site translations, with bond dimensions up to $m = 300$.
Once the wavefunction converges~\footnote{Our convergence criterion is based on the overlap between the optimized and predicted local wavefunctions as suggested in Ref.~\onlinecite{mcculloch2008infinite}. We assume that a wavefunction has converged when the overlap either becomes smaller than $10^{-14}$ or reaches the truncation error}, we compute the two-site transfer matrix $T$ [see Fig.~\ref{fig:iDMRG}(a)] and its eigenvalues $\lambda_i$ with $i\in {0, 1,2, \ldots}$. 
The largest eigenvalue $\lambda_0 = 1$ dictates the normalization condition for the wavefunction while subsequent ones characterize any correlations in the ground state.
In particular, the second-largest eigenvalue $\lambda_1$ encodes the length of the longest correlations and its wavenumber via the relation 
\begin{align}
\label{eq:T_eig}
    \lambda_1 = \exp{\left[2\left(-\frac{1}{\xi} + \mathrm{i}\, k \right)\right]}\,.
\end{align} 
This relation can be explicitly seen by considering the diagram in Fig.~\ref{fig:iDMRG}(b): this translates to any two-point correlator of the form $\mathcal{C}\, (2d)= \langle O_i O_{i+2d+1}\rangle = (L| \mathcal{O}\, T^d\, \mathcal{O}|R)$, where $(L|$ and $|R)$ are the left and right eigenvectors of $T$ for the largest eigenvalue $\lambda_0$, and $\mathcal{O}$ corresponds to the expectation value of an operator $O$ with respect to a local tensor in the MPS representation. Since any correlation decays as $\sim \lambda_i^{d}$, at sufficiently large distances, it is dominated by $\lambda_1$. 

Proceeding along these lines, Fig.~\ref{fig:iDMRG}(c) shows the variation of the correlation length $\xi$ with the parameter $f$ along the two specific cuts $L_1$ ($\theta = \pi/16$) and $L_2$ ($\theta =\pi/8$) marked in Fig.~\ref{fig:CCM_PD}. In consistency with our previous phase diagram, $\xi$ diverges near the expected critical points. A particularly interesting feature is captured by Fig.~\ref{fig:iDMRG}(d), which plots $2\,k$ (mod $2\pi$) against the inverse correlation length $2/\xi$. For both the lines $L_1$ and $L_2$, the wavevector $k$ remains nontrivial even close to the critical point, although it eventually approaches zero as the correlation length diverges. We fit this data to $k\sim (1/\xi)^\ell$ and find the best fits $\ell\approx 0.77$ ($\theta=\pi/16$) and $\ell\approx 0.64$ ($\theta =\pi/8$) for $\xi < 100$ [Fig.~\ref{fig:iDMRG}(d), inset], whereafter it appears that the effect of finite bond dimensions becomes significant. The variations in $\ell$ are likely a consequence of the proximity of the Potts critical point (which has $k=0$), and we do not appear to have captured the ultimate scaling behavior. 
Nonetheless, the implications of this variation in $k$ are twofold. First, they corroborate that the QPT from the disordered to the commensurate $\mathbb{Z}_3$-ordered phase is indeed a direct transition and Fig.~\ref{fig:iDMRG} succinctly highlights why this is possible: the period of the incommensuration ($\equiv p \sim 1/k$) diverges as the correlation length does. Secondly, these observations signify that the diverging correlations in the vicinity of the critical points may exhibit nontrivial spatial structures such as modulations.

\section{Rydberg array and constrained hard-boson model}
\label{sec:boson}
\noindent
As mentioned in Sec.~\ref{sec:intro}, the universal behavior of the $\mathbb{Z}_3$ CCM is also reflected in the phase diagram of an array of Rydberg atoms described by Eq.~\eqref{eq:Rydberg}. To see this, we consider the case when $V_1\gg |\Omega|,|\delta|$, that is, nearest-neighbor interactions are so strong that effectively no two neighboring atoms can simultaneously be in the Rydberg state. Further, owing to the rapidly decaying form  of the interactions $V_x =C_6/x^6$, we can approximate the Hamiltonian by neglecting couplings beyond the third-nearest neighbor: $V_x\approx 0$ for $x\geq 3$. With this truncation, we arrive at a toy model of the form 
\begin{align}\label{eq:Rydberg_toy}
H_{\rm Ryd}=&\sum_{i=1}^N \frac{\Omega}{2} (\ket{g}_i\!\bra{r}+\ket{r}_i\!\bra{g})-\delta \ket{r}_i\!\bra{r}\nonumber\\
&+V_{2}\ket{r}_i\!\bra{r}\otimes \ket{r}_{i+2}\!\bra{r}
\end{align}
with the constraint $\ket{r}_i\!\bra{r}\otimes \ket{r}_{i+1}\!\bra{r}=0$.
This model, which also appears in studies of ultracold atoms in tilted optical lattices \cite{greiner2002quantum}, has been analyzed in the literature in the context of interacting bosons \cite{sachdev2002mott, fendley2004competing, GSS18}. The mapping to bosons (with annihilation operators $b_i$ and number operators $n_i=b_i^\dag b_i$) is apparent on identifying the state where the atom at site $i$ is in the internal state $\ket{g}$ with a vacuum state of a bosonic mode, and the state with the atom in $\ket{r}$ with the presence of a boson. In the hard-core limit, where no more than one boson can occupy a single site, the above Hamiltonian can be rewritten as 
\cite{sachdev2002mott, fendley2004competing}
\begin{align}
\label{eq:UV}
H_b=\sum_{i=1}^{N} -w\,(b_i^\dag+b_i)+U n_i +Vn_i n_{i+2},
\end{align}
supplemented with the constraint \begin{align}\label{1-constraint}
n_i \, n_{i+1}=0.
\end{align} 
This condition, that prohibits two bosons from occupying neighboring sites, is referred to as the one-site blockade constraint.
In this work, we refer to this system as the U-V model for obvious reasons. In order to connect this to previous literature, it is useful to note the correspondence $\Omega=-2 w$, $\delta=-U$, and $V_2=V$. The equilibrium phase diagram of the Hamiltonian \eqref{eq:UV} was studied as a function of its parameters in Refs.~\onlinecite{fendley2004competing, GSS18}. Depending on the couplings, the ground state can exhibit several different kinds of order. For example, if the chemical potential favors creating particles ($U <0$) and there is an attractive (repulsive) next-nearest neighbor potential set by $V < 0$ ($V > 0$), the ground state maximizes the density of bosons by having a particle on every second (third) site and the consequent ordered phase spontaneously breaks a $\mathbb{Z}_2$ ($\mathbb{Z}_3$) translational symmetry. More careful considerations show that when the energies of the two kinds of ordered states are comparable, there also exists a gapless IC phase \cite{fendley2004competing}, similar to the CCM.

In the following discussion, we concentrate on few special points in this phase diagram. Expressly, we also examine the limiting case $V\rightarrow \infty$, which further implies that all states obey the two-site blockade constraint
\begin{align}
\label{2-constraint}
n_i\,n_{i+2} = 0.
\end{align}

\subsection{Phase diagram}
\subsubsection{Transition from disorder to $\mathbb{Z}_2$ order}
Let us momentarily review the physics when $V=0$. In this case, we expect a second-order QPT from the disordered phase to the $\mathbb{Z}_2$-ordered phase as $u = U/w$ is varied from positive to negative values. As sketched earlier, we determine the location of the  phase transition and its nature from the aforementioned FSS analysis by numerically computing the gap as a function of $u$ for varying $N$, upon which, we find $U_c/w=1.308$ and $z=1$. Similarly, from previous data collapse arguments, we know that on plotting $N^z\Delta$ against $N^{1/\nu}(u-u_c)$, the resulting curves should ideally merge into a single one for different system sizes; this occurs for a value of $\nu =1$. Ipso facto, this phase transition unequivocally belongs to the Ising universality class.

\begin{figure*}[htb]
\begin{center}
\includegraphics[width=\linewidth]{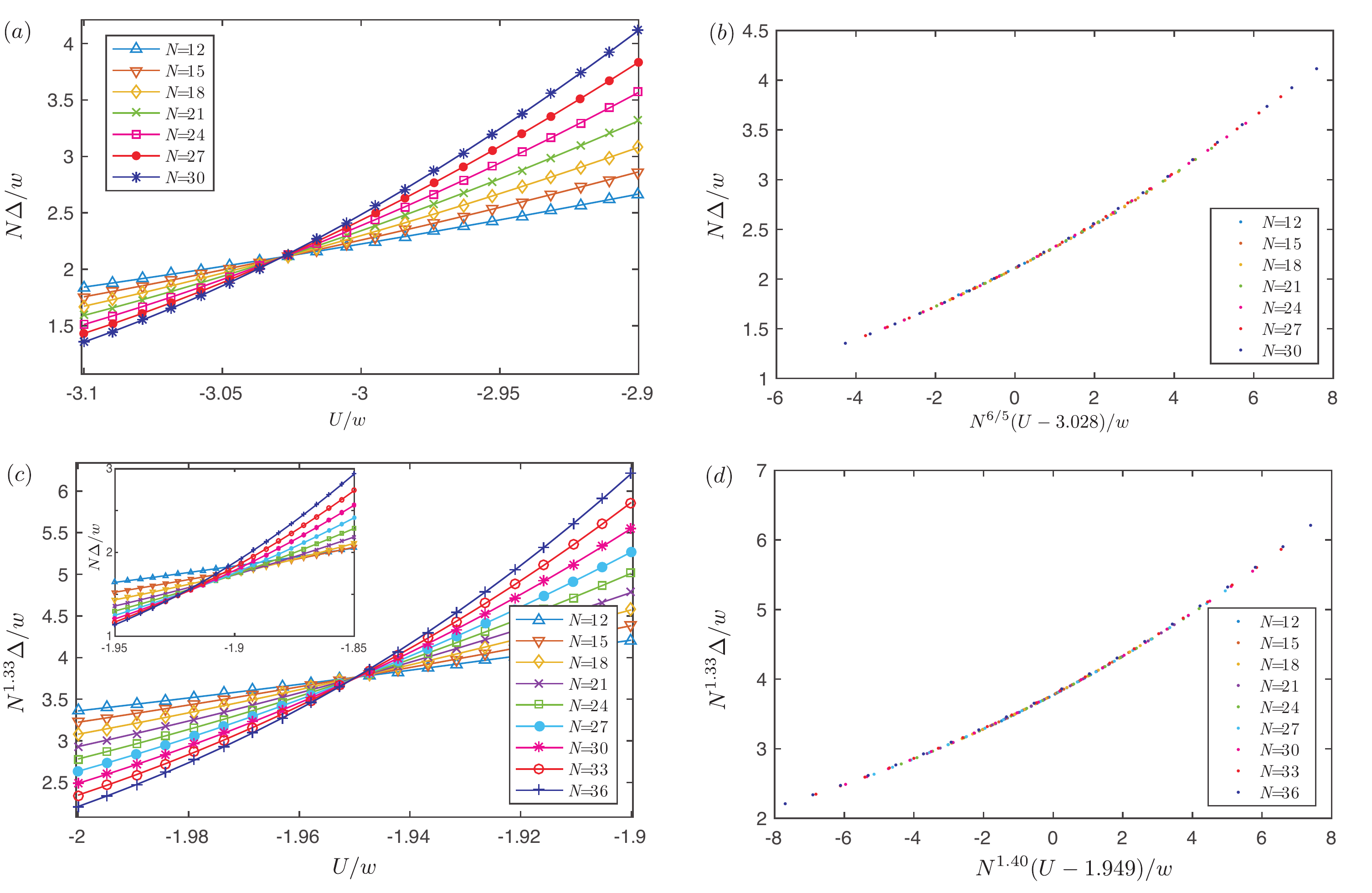}
\caption{\label{fig:Z3_R}[Upper panel]: FSS analysis of the QPT from the disordered to the $\mathbb{Z}_3$-ordered phase, across the integrable line $w^2=U\,V+V^2$, based on exact diagonalization of the Hamiltonian \eqref{eq:UV}. (a) Plot of the energy gap $\Delta$, rescaled by $N^{-z}$ with $z=1$, as a function of $U/w$ for different system sizes $N$. The sharp intersection of the different curves at $U/w=-3.029$ determines the QCP and conveys that  $z=1$. The location of the QCP agrees exactly with the analytically known result $U_c/w \approx -3.0299$ [Eq.~\eqref{eq:MCP}]. (b) Scaling plot of the energy gap for different system sizes; perfect data collapse is achieved for $\nu=5/6$. [Lower panel]: Same as above but in the limit $V \rightarrow \infty$. FSS now betokens $U/w=-1.949$, $z \approx 4/3$, and $\nu \approx 5/7$. }
\end{center}
\end{figure*}

\subsubsection{Transition from disorder to $\mathbb{Z}_3$ order}
\label{sec:exp_UV}

For finite $V/w >0$, we expect a QPT from the gapped disordered state, characterized by a featureless, low density state at large positive $U/w$, to the $\mathbb{Z}_3$-ordered phase, which is a density-wave state of period three at large negative $U/w$. However, as we demonstrate below, the transition to this phase with spontaneously broken $\mathbb{Z}_3$ symmetry can be of two fundamentally distinct types. This was originally argued for by \citet{fendley2004competing} in their proposal for a continuum quantum field theory description for the onset of Potts order. The most general effective action for such a theory, constructed in terms of a period-3 density-wave order parameter field, $\Psi$, is permitted by symmetry to include a linear derivative term $\mathrm{i}\,\alpha\,\Psi^*\,\partial_x \Psi$, which induces a chirality: as in Fig.~\ref{fig:Fig1}(f), with $\alpha \neq 0$, there are two inequivalent domain wall configurations. For $\alpha=0$, the transition is in the conformal three-state Potts class, as we demonstrate in Sec.~\ref{sec:pottsintegrable}. At the Potts point, $\alpha$ is a relevant perturbation \cite{cardy1993critical,fendley2004competing} with scaling dimension $1/5$. Consequently, for small nonzero $\alpha$, the transition turns into the direct chiral $\mathbb{Z}_3$ transition with $z \neq 1$, as we will see below. 

In the phase diagram of Fig.~1 of Ref.~\onlinecite{fendley2004competing}, a narrow IC phase was indicated immediately on one side of the Potts critical point, with a direct chiral $\mathbb{Z}_3$ transition between gapped phases on the other. We do not believe the IC phase extends all the way to the Potts point. For small $\alpha \neq 0$, the physics should be the same for both signs of $\alpha$ and hence, the Potts point should be flanked on both sides by direct chiral transitions from the gapped symmetric phase to a commensurate $\mathbb{Z}_3$-ordered phase. The latter scenario is the same as that in Fig.~2(a) of Ref.~\onlinecite{huse1982domain}, which indicated immediate direct chiral $\mathbb{Z}_3$ transitions on both sides of the Potts point.

\subsection{Critical exponents}
\subsubsection{Potts criticality on the integrable line}
\label{sec:pottsintegrable}

There are two special parameter-space lines in the phase diagram of the U-V model, defined by  
\begin{align}
w^2=U\,V+V^2,
\end{align}
along which the system is integrable \cite{fendley2004competing,baxter1980hard, baxter1982hard, baxter1983hard, huse1982tricriticality, baxter1981rogers, baxter1982exactly}. Along each of these, there is a quantum (multi)critical point at
\begin{align}
\label{eq:MCP}
\frac{V}{w}= \pm \bigg({\frac{\sqrt{5}+1}{2}}\bigg)^{5/2}.
\end{align}
Our interest lies in the point at positive $V/w$, which separates the disordered phase from the $\mathbb{Z}_3$-ordered phase, and is known to be a multicritical point in the universality class of the three-state Potts model. The same FSS arguments as above confirm the location of this multicritical point with very high accuracy and the critical exponents, moving across this integrable line, are found to be $z=1$ and $\nu=5/6$, as expected [see Fig.~\ref{fig:Z3_R}(a--b)].

\subsubsection{Chiral transition in the two-site blockade limit}
The transition away from the integrable line, however, does not belong to the Potts universality class. In this regard, the more interesting case that we now turn to is the limit $V\rightarrow \infty$, $\lvert U \rvert \ll V$, where the two-site blockade constraint is enforced. As depicted in Fig.~\ref{fig:Z3_R}, the scaling of the gap once again tells us about the critical point and exponents. With these parameters, we establish  $u_c=-1.949$, $z \approx 1.33$, and $1/\nu \approx 1.40$. Thus, we indeed recover (within numerical error estimates) the same critical exponents as for the chiral clock model in Sec.~\ref{sec:P0}. The discrepancies between the two are reasonable given that these models are equivalent only in the sense of universality (without the requirement of a one-to-one mapping of their Hilbert spaces).

\subsubsection{Finite-size scaling along the chiral phase boundary}
\begin{figure}[htb]
\begin{center}
\includegraphics[width=\linewidth]{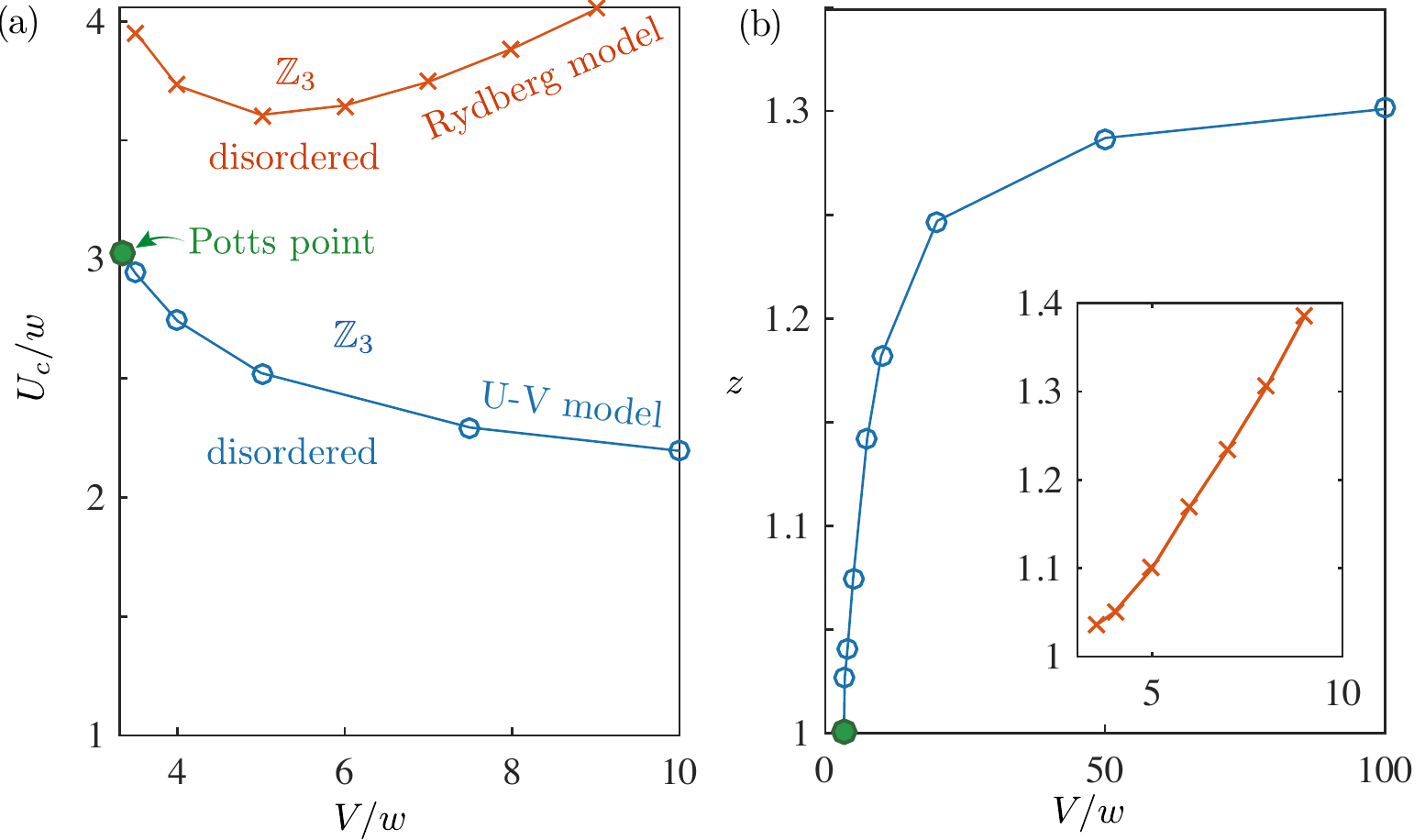}
\caption{(a) Phase boundary between the disordered and the $\mathbb{Z}_3$-ordered state above the Potts critical point [Eq.~(\ref{eq:MCP})] with nearest-neighbor/next-nearest-neighbor couplings only (U-V-model, blue) and long-range Rydberg interactions (red). In the latter case, the interaction strength between two spins separated by $x$ sites is given by $V_x =  C_6/x^6$. To plot the phase boundary on the same scale, we identify $ V_2= V$. (b) The dynamical critical exponent $z$ when crossing this phase boundary in the U-V model, obtained from FSS with up to $N=21$ sites (blue). At the Potts point (green), we find $z=1$ as expected. Away from the Potts point, $z$ increases and for $V\rightarrow \infty$, we reckon an asymptotic value of $z \approx 1.33$; the crossover in $z$ between these limits is likely a finite-size effect. The inset shows the analogously extracted value of $z$ for the Rydberg Hamiltonian. Here also, we find that $z$ increases with the interaction strength, starting from a value close to one---this is consistent with the presence of a Potts point in the Rydberg model as well. Note that in contrast to the U-V model, here, increasing $V$ would eventually lead to the appearance of the the $\mathbb{Z}_4$ phase, which is not discussed in this work. The FSS analysis for the Rydberg model is limited to comparatively smaller systems, of up to 18 atoms, owing to the sizeably larger Hilbert space.}\label{Fig4}
\end{center}
\end{figure}

One can iterate over the same procedure for all values of $V/w$ in the interval $\left((\sqrt{5}+1)/2)^{5/2},\infty\right]$ and numerically extract the critical parameters and exponents; the results of such a calculation are showcased in Fig.~\ref{Fig4}. In this fashion, we recover the schematic phase boundary proposed in Ref.~\onlinecite{fendley2004competing}. Akin to the CCM in Sec.~\ref{sec:P0}, the FSS analysis yields critical exponents that vary monotonically from $z\approx 1.33$ at $V\rightarrow \infty$ (i.e., far away from the Potts point) to 1 at the Potts critical point. 

\subsection{Effect of long-range interactions}
\label{sec:lr}
Unlike the idealized models \eqref{eq:Rydberg_toy} and \eqref{eq:UV} considered above, in realistic experiments with trapped Rydberg atoms \cite{jaksch2000fast, Browaeys09, weimer2010rydberg, labuhn2016tunable, bernien2017probing, kim2018detailed}, the interactions between excited states have nonzero tails, decaying as $V_x = C_6/x^6$. In order to probe the physics behind the QPTs in these systems, one needs to understand the influence of these long-range couplings---it is not \textit{a priori} clear whether these corrections to the interactions could change the nature of the transition or even the topology of the phase diagram.
Indeed, for sufficiently large $C_6$ and $ |\delta| \gg \Omega$, it has been previously suggested that the ground states of the Rydberg Hamiltonian $H_\textrm{Ryd}$ display a series of new phases with distinct spatial symmetry breaking \cite{sela2011dislocation, pohl2010dynamical} such as $\mathbb{Z}_4$, $\mathbb{Z}_5$ and the like, in which the Rydberg atoms are arranged regularly across every fourth or fifth site on the array; this is in contrast to both the CCM and U-V models.

Nevertheless, from numerical simulations combining exact diagonalization and FSS analogous to Sec.~\ref{sec:FSS}, we find that the system continues to exhibit a direct transition from the disordered to the $\mathbb{Z}_3$-ordered phase in certain parameter regimes. While both the Potts and chiral $\mathbb{Z}_3$ transitions still persist, the associated phase boundaries and critical exponents are nontrivially altered as can be seen in Fig.~\ref{Fig4}. In the presence of $C_6/x^6$ couplings, the critical points are shifted to larger values of $U$---this is understandable since the long tails tend to energetically favor the disordered state \cite{jaschke2017critical}. Note, however, that in the Rydberg model, reaching arbitrarily large values of $V/w$ becomes difficult due to the onset of the $\mathbb{Z}_4$-ordered phase, which has no counterpart in the U-V model. Likewise, the dynamical critical exponent $z$ is modified in that it attains its saturation value for smaller $V/w$. Qualitatively, this is because the long-range interactions enhance the inequivalence between the two kinds of domain walls (thereby rendering the system ``more chiral'' in some sense), which, crudely, translates to a faster deviation from the Potts exponent. Remarkably, these chiral critical exponents can actually be observed in the quench dynamics of ultracold atomic systems---such as the 51-atom Rydberg simulator---through the Kibble-Zurek mechanism \cite{kibble1976topology, kibble1980some, zurek1985cosmological, zurek1993cosmic, zurek1996cosmological}, which gives one access to the combination $\nu / (1+ z\,\nu)$. The detailed numerical \cite{pirvu2010matrix} and experimental study of this model will be presented in a future work \cite{keesling2018}.

\section{Conclusions}
\label{sec:end}

Motivated by recent experiments observing ``Rydberg crystals'' \cite{fendley2004competing, pohl2010dynamical} in a one-dimensional chain of ultracold atoms \cite{bernien2017probing}, we have examined the direct quantum phase transition (QPT) between a gapped phase with no symmetry breaking, and a $\mathbb{Z}_3$-ordered phase. Such a quantum phase transition is directly realized in the one-dimensional chiral clock model (CCM) for three-state spins, $H_\textsc{ccm}$, in Eq.~(\ref{eq:Hamiltonian}). One of the advantages of the CCM is that it allows us to numerically study fairly large system sizes. The exponents for this case are cataloged in Table~\ref{Table:z}. Once we move away from the achiral Potts transition at $\theta =0$, there is clear evidence for a dynamical critical exponent of value $z>1$. 

In the same vein, we also studied the lattice boson model, $H_b$, in Eq.~(\ref{eq:UV}) that was first proposed by \citet{fendley2004competing}. This is expected to display a lattice-translational-symmetry-breaking QPT in the same universality class as that in $H_\textsc{ccm}$ with the parameters $\phi=0$ and $\theta \neq 0$. On that account, we presented a numerical study of the critical properties of this model and obtained confirmation of a continuous phase transition with $z \approx 1.33$ and $\nu \approx 0.71$---these values are close to the exponents found for $H_\textsc{ccm}$. Taken together, our results manifestly imply the existence of a strongly coupled critical theory which is not a conformal field theory, or even relativistic.

Our results also elucidate how a direct transition to a $\mathbb{Z}_3$-ordered phase can occur in a chiral model, without an intermediate gapless incommensurate phase. With nonzero chirality in the Hamiltonian, correlations in the gapped disordered phase have an oscillatory character, which decays exponentially at long length scales. 
However, the period of the incommensurability diverges as the transition is approached: this is demonstrated by the results in Fig.~\ref{fig:iDMRG}. This complex behavior highlights the strong-coupling nature of the direct chiral $\mathbb{Z}_3$ quantum transition, and underlies the difficulty in obtaining a field-theoretic renormalization group description.

\begin{acknowledgments}

We acknowledge helpful discussions with Paul Fendley, Yin-Chen He, Jong Yeon Lee, Krishnendu Sengupta, and Seth Whitsitt.
This research was supported by the National Science Foundation (NSF) under Grants DMR-1664842, PHY-1506284, the Center for Ultracold Atoms (NSF Grant PHY-1734011), and the Vannevar Bush Faculty Fellowship. S.C. acknowledges support from the Kwanjeong Educational Foundation. H.P. is supported by the NSF through a grant at the Institute for Theoretical Atomic Molecular and Optical Physics (ITAMP) at Harvard University and the Smithsonian Astrophysical Observatory. The computations in this paper were run on the Odyssey cluster supported by the FAS Division of Science, Research Computing Group at Harvard University.  
\end{acknowledgments}

\bibliographystyle{apsrev4-1_custom}
\bibliography{Z3Ref.bib}
%\bibliography{HPref.bib}

\end{document}